\documentclass[aps,prd,superscriptaddress,nofootinbib,secnumarabic]{revtex4-2}[10pt]
\pdfoutput=1
\usepackage{latexsym}
\usepackage{mathrsfs}
\usepackage{slashed}
\usepackage{amsmath}
\usepackage{amssymb}
\usepackage{unicode}
\usepackage{diagbox}
\usepackage{makecell}
\usepackage{hyperref}
\usepackage{url}
\usepackage{graphicx}
\usepackage{appendix}
\usepackage{indentfirst}
\usepackage{layout}
\usepackage{geometry}

\begin{document}
\title{Higgs boson to $\gamma Z$ decay as a probe of flavour changing neutral Yukawa couplings}

\author{Shi-Ping He}
\email{sphe@ihep.ac.cn}
\affiliation{Center for Future High Energy Physics and Theoretical Physics Division, Institute of High Energy Physics, Chinese Academy of Sciences, Beijing 100049, China}

\date{\today}

\begin{abstract}
With the deeper study of Higgs particle, Higgs precision measurements can be served to probe new physics indirectly. In many new physics models, vector-like quarks $T_L,T_R$ occur naturally. It is important to probe their couplings with standard model particles. In this work, we consider the singlet $T_L,T_R$ extended models and show how to constrain the $Tth$ couplings through the $h\rightarrow\gamma Z$ decay at high-luminosity LHC. Firstly, we derive the perturbative unitarity bounds on $|y_{L,R}^{tT}|$ with other couplings set to be zeros simply. To optimize the situation, we take $m_T=400$ GeV and $s_L=0.2$ considering the experimental constraints. Under this benchmark point, we find that the future bounds from $h\rightarrow\gamma Z$ decay can limit the real parts of $y_{L,R}^{tT}$ in the positive direction to be $\mathrm{O}(1)$ because of the double enhancement. For the real parts of $y_{L,R}^{tT}$ in the negative direction, it is always surpassed by the perturbative unitarity. Moreover, we find that the top quark electric dipole moment can give stronger bounds (especially the imaginary parts of $y_{L,R}^{tT}$) than the perturbative unitarity and $h\rightarrow\gamma Z$ decay in the off-axis regions for some scenarios.
\end{abstract}

\maketitle

\clearpage
\section{Introduction}
The standard model (SM) of elementary particle physics was proposed in the 1960s \cite{Glashow:1961tr, *Weinberg:1967tq, *Salam:1968rm}, and it has been verified to be quite successful up to now. However, there are still many problems beyond the ability of SM, for example, Higgs mass naturalness, gauge coupling unification, fermion mass hierarchy, electro-weak vacuum stability, dark matter, matter anti-matter asymmetry, and so on. Thus, new physics beyond the SM (BSM) are motivated in the high energy physics community. Many of these BSM models predict the existence of heavy fermions, for example, composite Higgs models \cite{Agashe:2004rs, Panico:2015jxa}, little Higgs models \cite{ArkaniHamed:2002qy, Schmaltz:2005ky}, grand unified theories \cite{Hewett:1988xc}, extra dimension models
\cite{Contino:2006nn}. In these models, there can be a heavy up-type quark $T$, which interacts with the SM particles through $TbW,TtZ,Tth$ interactions. Analyses on these couplings may tell us some clues about the new physics. $TbW$ coupling can be constrained from single production of $T$ quark, but there are always many assumptions for most of the current constraints. It will be hard for the detection of the flavour changing neutral (FCN) couplings $TtZ,Tth$, because $T$ productions from $tZ,th$ fusion are highly suppressed. If there exist other new decay channels for the $T$ quark \footnote{Say $T\rightarrow tS$, here $S$ can be a $CP$ even or odd new scalar.}, even the bounded $TbW$ coupling can be saturated. Since the discovery of Higgs boson at the Large Hadron Collider (LHC) \cite{Aad:2012tfa, *Chatrchyan:2012xdj}, it can also be a probe to such new physics.

Currently, all the main production and decay channels of the Higgs boson have been discovered at the LHC. Then, the next step is to measure the observed channels more accurately. At the same time, attention should be paid to the undiscovered channels. Precision measurements of the Higgs particle may help us decipher the nature of electro-weak symmetry breaking (EWSB) \cite{Englert:1964et, *Higgs:1964ia, *Guralnik:1964eu, *Kibble:1967sv} and open the door to new physics \cite{Dittmaier:2011ti, *Dittmaier:2012vm, *Heinemeyer:2013tqa, *deFlorian:2016spz}. $hVV$ and $hff$ couplings inferred from the observed channels are SM-like now, while there can still be large deviations for the rare decay modes, for example, $h\rightarrow\gamma Z,\mu^+\mu^-$. The $\gamma Z$ decay mode has drawn much attention of this community. It can be used to detect $CP$ violation \cite{Chen:2014ona, Farina:2015dua, Chen:2017plj} and many new physics scenarios \cite{Chen:2013vi, No:2016ezr, Dawson:2018pyl}. Here, we will show how to constrain the FCN Yukawa (FCNY) couplings through the $h\rightarrow\gamma Z$ decay mode indirectly. The constraints from $h\rightarrow\gamma Z$ do not depend on the total width of $T$ quark, namely, in spite of other decay modes.

In this paper, we build the framework of FCN couplings in Sec.~\ref{sec:framework} firstly. Sec.~\ref{sec:constraints} is devoted to the theoretical and experimental constraints on the simplified model. In Sec.~\ref{sec:width}, we compute the new physics contributions to the partial decay width of $h\rightarrow\gamma Z$. Then, we perform the numerical constraints on the FCNY interactions in Sec.~\ref{sec:numerical}. Finally, we give the summary and conclusions in Sec.~\ref{sec:summary}.
\section{Framework of flavour changing neutral couplings}\label{sec:framework}
\subsection{UV complete model}
It is strongly constrained for the mixings between heavy particles and the first two generations because of the bounds from flavour physics \cite{delAguila:2000aa, delAguila:2000rc, AguilarSaavedra:2002kr}. What is more, the third generation is more likely to be concerned with new physics theoretically owing to the mass hierarchy. For convenience and simplicity, we only take into account the mixings between the third generation and heavy quarks.

Based on the SM gauge group $SU_C(3)\otimes SU_L(2)\otimes U_Y(1)$, we can enlarge the SM by adding new particles with different representations. Usually, we extend the SM fermion sector by introducing vector-like particles to avoid the quantum anomaly. The minimal extension of quark sector is to add one pair of vector-like quarks (VLQs) \cite{AguilarSaavedra:2009es, Aguilar-Saavedra:2013qpa}. For the non-minimally extended models, the scalar sector can also be augmented. Besides the VLQs, we can also plus a real gauge singlet scalar \cite{He:2014ora, Dolan:2016eki, Kim:2018mks}, a Higgs doublet \cite{Aguilar-Saavedra:2017giu}, and even both the singlet-doublet scalars at the same time \cite{Muhlleitner:2016mzt, Aguilar-Saavedra:2017giu}. In these models, there can exist other decay channels \cite{Cheung:2018ljx, Cacciapaglia:2019zmj}.

FCN couplings $TtZ$ and $Tth$ show different patterns in different models. For simplicity, we will only consider the case where there is one pair of VLQs $T_L$ and $T_R$. \footnote{Of course, one can build one model with more $T_L,T_R$ quarks. But the mass matrix may be equal to and even greater than three dimensions, which are quite complex.}. In the following, we will give two specific examples: the minimal extension with a pair of singlet quarks $T_L,T_R$ (gauge group representation (3, 0, 2/3)) and the model further enlarged by an extra real singlet scalar (gauge group representation (1, 0, 0)).
\subsubsection{Minimal vector-like quark model}
Let us start with the model by adding a pair of singlet fermions $T_L,T_R$ to the SM, which is dubbed as the VLQT model. The Lagrangian can be written as \cite{Aguilar-Saavedra:2013qpa}
\begin{align}
&\mathcal{L}=\mathcal{L}_{SM}+\mathcal{L}_T^{Yukawa}+\mathcal{L}_T^{gauge},\nonumber\\
&\mathcal{L}_T^{Yukawa}=-\Gamma_T^i\bar{Q}_L^i\widetilde{\Phi}T_R-M_T\bar{T}_LT_R+\mathrm{h.c.},\quad\mathcal{L}_T^{gauge}=\bar{T}_Li\slash\!\!\!\!DT_L+\bar{T}_Ri\slash\!\!\!\!DT_R,
\end{align}
where $\widetilde{\Phi}=i\sigma_2\Phi^{\ast}$ and the covariant derivative is defined as $D_{\mu}=\partial_{\mu}-ig^{\prime}Y_TB_{\mu}$. $Y_T$ and $Q_T$ are the $U_Y(1)$ and electric charge of the $T$ quark, respectively. The Higgs doublet is parametrized as $\Phi^T=[\phi^+,~\frac{v+h+i\chi}{\sqrt{2}}]$.

It is easy to obtain the mass terms of $t$ and $T$ quarks \footnote{Although the mass mixing term $\bar{T}_Lt_R$ can appear, it will be removed via field redefinition \cite{Dawson:2012di, DeSimone:2012fs}.}:
\begin{align}\label{eqn:quark:mass}
\mathcal{L}_{mass}\supset-
\left[\begin{array}{cc} \bar{t}_L&\bar{T}_L \end{array}\right]
\left[\begin{array}{cc}\frac{1}{\sqrt{2}}\Gamma_t^{33}v&\frac{1}{\sqrt{2}}\Gamma_T^3v\\ 0 &M_T\end{array}\right]
\left[\begin{array}{cc} t_R\\T_R \end{array}\right]+\mathrm{h.c.}~.
\end{align}
Here, $\Gamma_T^3$ and $\Gamma_t^{33}$ are the gauge eigenstate Yukawa couplings in front of $\bar{Q}_L^3\widetilde{\Phi}T_R$ and $\bar{Q}_L^3\widetilde{\Phi}t_R$ individually. The $t$ and $T$ quarks can be rotated into mass eigenstates by the following transformations:
\begin{align}\label{eqn:quark:rotation}
\left[\begin{array}{c}t_L\\T_L\end{array}\right]\rightarrow
	\left[\begin{array}{cc}\cos\theta_L&\sin\theta_L\\-\sin\theta_L&\cos\theta_L\end{array}\right]
	\left[\begin{array}{c}t_L\\T_L\end{array}\right],\quad
\left[\begin{array}{c}t_R\\T_R\end{array}\right]\rightarrow
	\left[\begin{array}{cc}\cos\theta_R&\sin\theta_R\\-\sin\theta_R&\cos\theta_R\end{array}\right]
	\left[\begin{array}{c}t_R\\T_R\end{array}\right].
\end{align}
Then, we have the following mass eigenstate Yukawa interactions:
\begin{align}
&\mathcal{L}_{Yukawa}\supset-m_t\bar{t}t-m_T\bar{T}T-\frac{m_t}{v}c_L^2h\bar{t}t-\frac{m_T}{v}s_L^2h\bar{T}T\nonumber\\
&-\frac{m_T}{v}s_Lc_Lh(\bar{t}_LT_R+\bar{T}_Rt_L)-\frac{m_t}{v}s_Lc_Lh(\bar{T}_Lt_R+\bar{t}_RT_L).
\end{align}
Here, $s_L,c_L,s_R,c_R$ are short for $\sin\theta_L,\cos\theta_L,\sin\theta_R,\cos\theta_R$, respectively. Similarly, we abbreviate $\sin\theta,\cos\theta$ as $s_{\theta},c_{\theta}$ in the following context. In this model, we have two independent extra parameters $m_T$ and $\theta_L$. There are two relations between the mixing angles and $t,T$ quark masses:
\begin{align}
\tan\theta_R=\frac{m_t}{m_T}\tan\theta_L,~M_T^2=m_T^2c_L^2+m_t^2s_L^2.
\end{align}

For the singlet $T_L$ and $T_R$ quarks, the gauge eigenstate $t,T$ quarks will interact with $Z,W$ bosons through the following form:
\begin{align}\label{eqn:quark:gauge}
&\mathcal{L}_{gauge}\supset\frac{g}{c_W}Z_{\mu}[\bar{t}_L\gamma^{\mu}(\frac{1}{2}-\frac{2}{3}s_W^2)t_L-\frac{2}{3}s_W^2\bar{t}_R\gamma^{\mu}t_R-Q_Ts_W^2(\bar{T}_L\gamma^{\mu}T_L+\bar{T}_R\gamma^{\mu}T_R)]\nonumber\\
&\qquad\qquad\qquad\qquad+\frac{g}{\sqrt{2}}(W_{\mu}^+\bar{t}_L\gamma^{\mu}b_L+W_{\mu}^-\bar{b}_L\gamma^{\mu}t_L).
\end{align}
Here, $t$ and $T$ quarks can be rotated into mass eigenstates by the transformations in Eq.~\eqref{eqn:quark:rotation}. Thus, we have the following mass eigenstate gauge interactions:
\begin{align}\label{eqn:quark:gauge0}
&\mathcal{L}_{gauge}\supset\frac{g}{c_W}Z_{\mu}[(\frac{1}{2}c_L^2-\frac{2}{3}s_W^2)\bar{t}_L\gamma^{\mu}t_L+(\frac{1}{2}s_L^2-\frac{2}{3}s_W^2)\bar{T}_L\gamma^{\mu}T_L+\frac{1}{2}s_Lc_L(\bar{t}_L\gamma^{\mu}T_L+\bar{T}_L\gamma^{\mu}t_L)\nonumber\\
&-\frac{2}{3}s_W^2\bar{t}_R\gamma^{\mu}t_R-\frac{2}{3}s_W^2\bar{T}_R\gamma^{\mu}T_R]+\frac{gc_L}{\sqrt{2}}(W_{\mu}^+\bar{t}_L\gamma^{\mu}b_L+W_{\mu}^-\bar{b}_L\gamma^{\mu}t_L)+\frac{gs_L}{\sqrt{2}}(W_{\mu}^+\bar{T}_L\gamma^{\mu}b_L+W_{\mu}^-\bar{b}_L\gamma^{\mu}T_L).
\end{align}
\subsubsection{Vector-like quark and one singlet scalar model}
Now, let us consider the model with SM extended by a pair of singlet VLQs $T_L,T_R$ and a real singlet scalar $S$, which is named as VLQT+S. The Lagrangian can be written as \cite{Dolan:2016eki, Xiao:2014kba}
\begin{align}
&\mathcal{L}=\mathcal{L}_{SM}+\mathcal{L}_T^{Yukawa}+\mathcal{L}_T^{gauge}+\mathcal{L}_S,\nonumber\\
&\mathcal{L}_T^{Yukawa}=-\Gamma_T^i\bar{Q}_L^i\widetilde{\Phi}T_R-M_T\bar{T}_LT_R-y_T^SS\bar{T}_LT_R+\mathrm{h.c.},\quad\mathcal{L}_T^{gauge}=\bar{T}_Li\slash\!\!\!\!DT_L+\bar{T}_Ri\slash\!\!\!\!DT_R,\nonumber\\
&\mathcal{L}_S=\frac{1}{2}\partial_{\mu}S\partial^{\mu}S-V_{\Phi S},\quad V_{\Phi S}=\mu_{\Phi{S}}\Phi^{\dag}\Phi{S}+\lambda_{\Phi{S}}
\Phi^{\dag}\Phi{S^2}+t_SS+m_S^2S^2+\mu_SS^3+\lambda_SS^4.
\end{align}
Note that the Lagrangian form is invariant after shifting $S$; thus, we can set $\langle S\rangle=0$ through redefinition of the scalar field $S$ \cite{Barger:2007im, Chen:2014ask, Kanemura:2015fra, He:2016sqr, Kanemura:2016lkz, Lewis:2017dme}. Here, $h$ can mix with $S$, so we should transform them into mass eigenstates via following rotations:
\begin{align}
\left[\begin{array}{c}h\\S\end{array}\right]\rightarrow
	\left[\begin{array}{cc}\cos\theta&\sin\theta\\-\sin\theta&\cos\theta\end{array}\right]
	\left[\begin{array}{c}h\\S\end{array}\right].
\end{align} 
The mass terms of $t$ and $T$ quarks are exactly the same as those in Eq.~\eqref{eqn:quark:mass}; thus, they can be rotated into mass eigenstates by the same transformations of Eq.~\eqref{eqn:quark:rotation}. There is one extra Yukawa term $-y_T^SS\bar{T}_LT_R$ compared to the model VLQ, so the Yukawa interactions in this model are more complex. Then we have the following mass eigenstate Yukawa interactions:
\begin{align}
&\mathcal{L}_{Yukawa}\supset-m_t\bar{t}t-m_T\bar{T}T-[\frac{m_t}{v}c_L^2c_{\theta}-\mathrm{Re}(y_T^S)s_Ls_Rs_{\theta}]h\bar{t}t-[\frac{m_T}{v}s_L^2c_{\theta}-\mathrm{Re}(y_T^S)c_Lc_Rs_{\theta}]h\bar{T}T\nonumber\\
&-[\frac{m_T}{v}s_Lc_Lc_{\theta}+\mathrm{Re}(y_T^S)s_Lc_Rs_{\theta}]h(\bar{t}_LT_R+\bar{T}_Rt_L)-[\frac{m_t}{v}s_Lc_Lc_{\theta}+\mathrm{Re}(y_T^S)c_Ls_Rs_{\theta}]h(\bar{T}_Lt_R+\bar{t}_RT_L).\nonumber\\
&+i\mathrm{Im}(y_T^S)s_Ls_Rs_{\theta}h\bar{t}\gamma^5t+i\mathrm{Im}(y_T^S)c_Lc_Rs_{\theta}h\bar{T}\gamma^5T\nonumber\\
&-i\mathrm{Im}(y_T^S)s_Lc_Rs_{\theta}h(\bar{t}_LT_R-\bar{T}_Rt_L)-i\mathrm{Im}(y_T^S)c_Ls_Rs_{\theta}h(\bar{T}_Lt_R-\bar{t}_RT_L).
\end{align}
The gauge interactions for $t$ and $T$ quarks are fully the same as those in Eq.~\eqref{eqn:quark:gauge} and Eq.~\eqref{eqn:quark:gauge0}. In this model, there are four interesting parameters $\theta_L,m_T,\theta,y_T^S$. The other parameters in scalar potential do not have relation with the $h\rightarrow\gamma Z,\gamma \gamma $ processes.
\subsection{Simplified model}
Here, we will adopt one more general and model independent framework \cite{Buchkremer:2013bha}. For simplicity, we only consider the singlet $T_{L,R}$ case. In the simplified model case, we can write down the related mass eigenstate state interactions generally:
\begin{align}\label{eqn:frame:simplify}
&\mathcal{L}\supset-m_t\bar{t}t-m_T\bar{T}T-eA_{\mu}\sum_{f=t,T}Q_f\bar{f}\gamma^{\mu}f+eZ_{\mu}[\bar{t}\gamma^{\mu}(g_L^t\omega_-+g_R^t\omega_+)t+\bar{T}\gamma^{\mu}(g_L^T\omega_-+g_R^T\omega_+)T\nonumber\\
&+\bar{t}\gamma^{\mu}(g_L^{tT}\omega_-+g_R^{tT}\omega_+)T+\bar{T}\gamma^{\mu}(g_L^{tT}\omega_-+g_R^{tT}\omega_+)t]-\frac{m_t}{v}h\bar{t}(\kappa_t+i\gamma^5\widetilde{\kappa}_t)t+h\bar{T}(y_T+i\gamma^5\widetilde{y}_T)T\nonumber\\
&+h\bar{t}(y_L^{tT}\omega_-+y_R^{tT}\omega_+)T+h\bar{T}((y_L^{tT})^{*}\omega_++(y_R^{tT})^{*}\omega_-)t+\frac{gc_L}{\sqrt{2}}(W_{\mu}^+\bar{t}_L\gamma^{\mu}b_L+W_{\mu}^-\bar{b}_L\gamma^{\mu}t_L)\nonumber\\
&+\frac{gs_L}{\sqrt{2}}(W_{\mu}^+\bar{T}_L\gamma^{\mu}b_L+W_{\mu}^-\bar{b}_L\gamma^{\mu}T_L),
\end{align}
where $\omega_{\pm}$ are the chirality projection operators $(1\pm\gamma^5)/2$ and the gauge couplings are listed as follows:
\begin{align}
&g_L^t=\frac{1}{s_Wc_W}(\frac{1}{2}c_L^2-\frac{2}{3}s_W^2),~g_L^T=\frac{1}{s_Wc_W}(\frac{1}{2}s_L^2-\frac{2}{3}s_W^2),~g_L^{tT}=\frac{s_Lc_L}{2s_Wc_W},\nonumber\\
&\qquad\qquad~~~g_R^t=-\frac{2s_W}{c_W},~g_R^T=-\frac{2s_W}{c_W},~g_R^{tT}=0.
\end{align}
Here, $m_T,\theta_L,\theta_R,\kappa_t,\widetilde{\kappa}_t,y_T,\widetilde{y}_T$ are all real parameters \footnote{Just as that we show in the above two models, $\theta_L,\theta_R$ may be not independent rotation angles in specific models \cite{Aguilar-Saavedra:2013qpa}.}, while $y_L^{tT},y_R^{tT}$ can be complex numbers. From now on, we will turn off the parameters $\widetilde{\kappa}_t$ and $\widetilde{y}_T$ for simplicity.

\begin{sloppypar}
\begin{table}[!h]
\begin{tabular}{c|c|c|c}
\hline
\diagbox{\qquad}{\qquad}& SM & VLQT & VLQT+S\\
\hline
$\kappa_t$ & 1 & $c_L^2$ & $c_L^2c_{\theta}-\frac{vs_Ls_Rs_{\theta}}{m_t}\mathrm{Re}(y_T^S)$\\
\hline
$y_T$ & $\times$ & $-\frac{m_T}{v}s_L^2$ & $-\frac{m_T}{v}s_L^2c_{\theta}+\mathrm{Re}(y_T^S)c_Lc_Rs_{\theta}$\\
\hline
$\mathrm{Re}(y_L^{tT})$ & $\times$ & $-\frac{m_t}{v}s_Lc_L$ & $-\frac{m_t}{v}s_Lc_Lc_{\theta}-\mathrm{Re}(y_T^S)c_Ls_Rs_{\theta}$\\
\hline
$\mathrm{Re}(y_R^{tT})$ &$\times$ & $-\frac{m_T}{v}s_Lc_L$ & $-\frac{m_T}{v}s_Lc_Lc_{\theta}-\mathrm{Re}(y_T^S)s_Lc_Rs_{\theta}$\\
\hline
$\mathrm{Im}(y_L^{tT})$ & $\times$ & 0 & $\mathrm{Im}(y_T^S)c_Ls_Rs_{\theta}$\\
\hline
$\mathrm{Im}(y_R^{tT})$ & $\times$ & 0 & $-\mathrm{Im}(y_T^S)s_Lc_Rs_{\theta}$\\
\hline
\end{tabular}
\caption{Patterns of Yukawa coefficients in SM, VLQT, and VLQT+S. There is no $T$ quark in SM, so we use the symbol $\times$ for $T$ couplings.} \label{tab:models:Yukawa}
\end{table}
\end{sloppypar}

\begin{sloppypar}
\begin{table}[!h]
\begin{tabular}{c|c|c|c|c}
\hline
\diagbox{\qquad}{\qquad}& $g_{L}^{tT}(y_L^{tT})^*$ & $g_{R}^{tT}(y_R^{tT})^*$ & $g_{L}^{tT}(y_R^{tT})^*$ & $g_{R}^{tT}(y_L^{tT})^*$\\
\hline
General & $\frac{1}{2s_Wc_W}s_Lc_L(y_L^{tT})^*$ & 0 & $\frac{1}{2s_Wc_W}s_Lc_L(y_R^{tT})^*$ & 0\\
\hline
VLQT & $-\frac{m_t}{v}\frac{s_L^2c_L^2}{2s_Wc_W}$ & 0 & $-\frac{m_T}{v}\frac{s_L^2c_L^2}{2s_Wc_W}$ & 0\\
\hline
VLQT+S & $\frac{s_Lc_L^2}{2s_Wc_W}(-\frac{m_t}{v}s_Lc_{\theta}-y_T^Ss_Rs_{\theta})$ & 0 & $\frac{s_L^2c_L}{2s_Wc_W}[-\frac{m_T}{v}c_Lc_{\theta}-(y_T^S)^*c_Rs_{\theta}]$ & 0\\
\hline
\end{tabular}
\caption{Patterns of the multiplication of FCN couplings in VLQT and VLQT+S.} \label{tab:models:gy}
\end{table}
\end{sloppypar}
In Tab.~\ref{tab:models:Yukawa}, we give the expressions of $\kappa_t,y_T,y_L^{tT},y_R^{tT}$ in three models. In Tab.~\ref{tab:models:gy}, we also give the expressions of $g_{L}^{tT}(y_L^{tT})^*,g_{R}^{tT}(y_R^{tT})^*,g_{L}^{tT}(y_R^{tT})^*,g_{R}^{tT}(y_L^{tT})^*$ in the VLQT and VLQT+S models. Here, we want to show the feasibility to constrain the FCNY couplings through the $h\rightarrow\gamma Z$ decay channel; thus, it is better to avoid drowning in elaborate theoretical calculations and collider phenomenology details. Although the FCNY couplings $y_{L,R}^{tT}$ are not free parameters in the mentioned VLQT and VLQT+S models, they can be free in more complex models because of enough degrees of freedom. For example, we can extend the SM by one pair of VLQs $T_L,T_R$ and many real singlet scalars. Here, we want to make a general analysis naively, so we take them to be free.
\section{Constraints on the model}\label{sec:constraints}
\subsection{Perturbative unitarity bound}
From theoretical point of view, large couplings may cause the problem of perturbative unitarity violation. One famous example is the upper limit of Higgs self-coupling (or the Higgs mass) in the SM \cite{Lee:1977eg}. For the scattering amplitude, we can perform the partial wave expansion:
$M=16\pi\sum\limits_{l=0}^{\infty}(2l+1)a_l(s)P_l(\cos\theta)$. Then, the partial $l$-wave amplitude is $a_l(s)=\frac{1}{32\pi}\int_{-1}^1d(\cos\theta)P_l(\cos\theta)M$. Especially, we have $a_0(s)=\frac{1}{32\pi}\int_{-1}^1d(\cos\theta)M$, which should satisfy $|\mathrm{Re}(a_0)|\leq\frac{1}{2}$. When you consider the two-to-two Higgs and longitudinally polarized vector boson scattering processes, $S$-wave unitarity will lead to the bound.

\begin{figure}[!h]
\begin{center}
\includegraphics[scale=0.3]{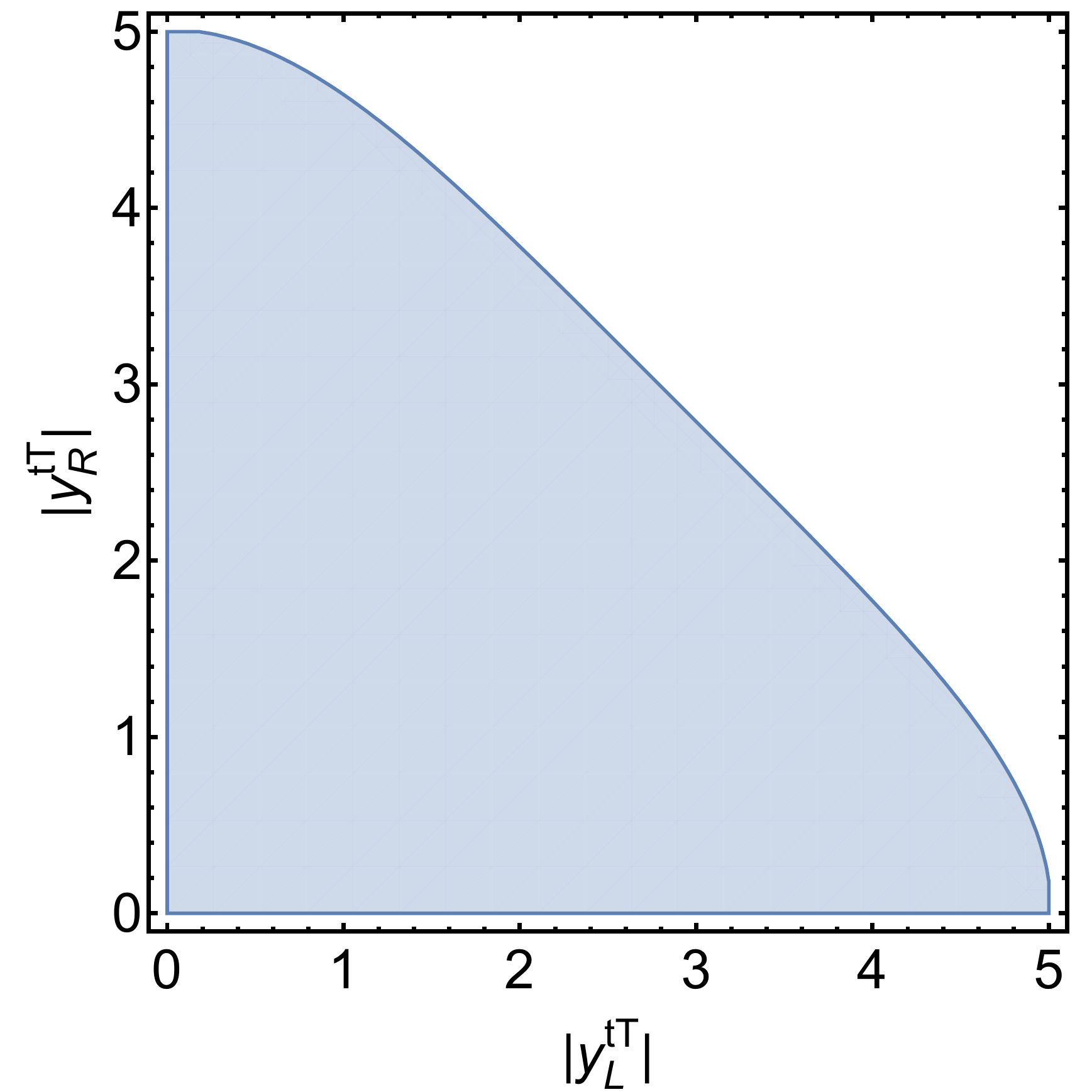}
\caption{The allowed region from perturbative unitarity in the plane of $|y_L^{tT}|-|y_R^{tT}|$.}\label{fig:unitarity}
\end{center}
\end{figure}

Similarly, we can bound the Yukawa couplings $y_{L,R}^{tT}$ from fermion scattering \cite{Chanowitz:1978mv, Maltoni:2001dc, Dicus:2004rg, Dicus:2005ku}. Then, we need to consider the two-to-two scattering processes with fermions. Obviously, there are two kinds of fermion processes: two-fermion and four-fermion processes. Actually, we only need to consider the neutral initial and final states. To simplify the analysis, we keep the $y_L^{tT},y_R^{tT}$ couplings but turn off the other couplings. After tedious computations, we get the following constraints (more details are given in App.~\ref{app:unitary}):
\begin{align}\label{eqn:unitarity}
\sqrt{(|y_L^{tT}|^2+|y_R^{tT}|^2)^2+12|y_L^{tT}|^2|y_R^{tT}|^2}+|y_L^{tT}|^2+|y_R^{tT}|^2\leq16\pi.
\end{align}
In Fig.~\ref{fig:unitarity}, we plot the parameter space region allowed by Eq.~\eqref{eqn:unitarity}.
\subsection{Constraints from direct search}
In the minimal extensions, the decay final states of $T$ are $bW^+,tZ,th$.
According to the Goldstone boson equivalence theorem, the partial decay widths satisfy the identity $\Gamma(T\rightarrow tZ)\approx\Gamma(T\rightarrow th)\approx\frac{1}{2}\Gamma(T\rightarrow bW)$ approximately (or $Br(T\rightarrow tZ)\approx Br(T\rightarrow th)\approx25\%,Br(T\rightarrow bW)\approx50\%$). For the pair production of VLQs, the cross section is determined by the strong interaction. It will give us the model independent bound on the $T$ quark mass, but we cannot get the information of $T$ quark couplings. Assuming $Br(T\rightarrow tZ)+Br(T\rightarrow th)+Br(T\rightarrow bW)=1$, the $T$ quark mass below 700 GeV$\sim$TeV is excluded at $95\%$ confidence level (CL) \cite{Aaboud:2018pii, Sirunyan:2019sza}. The $T$ quark can also be singly produced through $TbW$ coupling. In the singlet $T$ quark case, current constraints are $|s_L|\leq0.2$\cite{Aaboud:2018ifs}.

Current experiments give strong constraints on minimal VLQ models, but it will be relaxed in models with additional states. The mass can be light as 400 GeV if there exist an additional state mediated decay channels \cite{Cacciapaglia:2019zmj}. For the more complicated flavour and scalar sectors, there can be more than one mixing angle. The mixing angle is allowed to be larger.
\subsection{Constraints from electro-weak precision measurements}
The singlet VLQ $T_L,T_R$ will contribute to the $S,T$ parameters \cite{Peskin:1990zt, Peskin:1991sw}. The oblique corrections are mainly from the modification of SM gauge couplings and new particle loops. Their analytical expressions have been calculated in previous studies \cite{Lavoura:1992np, AguilarSaavedra:2002kr, Chen:2017hak}:
\begin{align}
\Delta S&\equiv S-S^{SM}\nonumber\\
	&=-\frac{N_t^Cs_L^2}{18\pi}[-2\log{r}_{tT}+c_L^2\frac{5-22r_{tT}^2+5r_{tT}^4}{(1-r_{tT}^2)^2}+c_L^2\frac{6(1+r_{tT}^2)(1-4r_{tT}^2+r_{tT}^4)}{(1-r_{tT}^2)^3}\log{r}_{tT}],\nonumber\\
\Delta T&\equiv T-T^{SM}=\frac{N_t^Cm_t^2s_L^2}{16\pi s_W^2m_W^2}(-1-c_L^2+\frac{s_L^2}{r_{tT}^2}-\frac{4c_L^2}{1-r_{tT}^2}\log{r}_{tT}),
\end{align}
with $r_{tT}\equiv\frac{m_t}{m_T}$. Now, let us define the $\Delta\chi^2$ as
\begin{align}
&\Delta\chi^2\equiv\sum\limits_{i,j=1,2}(O_i-O_i^{exp})(\sigma^2)_{ij}^{-1}(O_j-O_j^{exp}),
\end{align}
where $O_i\in\{\Delta S,\Delta T\},(\sigma^2)_{ij}=\sigma_i\rho_{ij}\sigma_j$. Their values are listed as follows \cite{Tanabashi:2018oca}:
\begin{align}
&\Delta S^{exp}=0.02,~\sigma_{\Delta S}=0.07,~\Delta T^{exp}=0.06,~\sigma_{\Delta T}=0.06,\nonumber\\
&\rho=\left[\begin{array}{cc}
			1&0.92\\0.92&1
			\end{array}\right],~
 \sigma^2=\left[\begin{array}{cc}
			\sigma_{\Delta S}&0\\0&\sigma_{\Delta T}
			\end{array}\right]\rho
			\left[\begin{array}{cc}
			\sigma_{\Delta S}&0\\0&\sigma_{\Delta T}
			\end{array}\right].
\end{align}
\begin{figure}[!h]
\begin{center}
\includegraphics[scale=0.35]{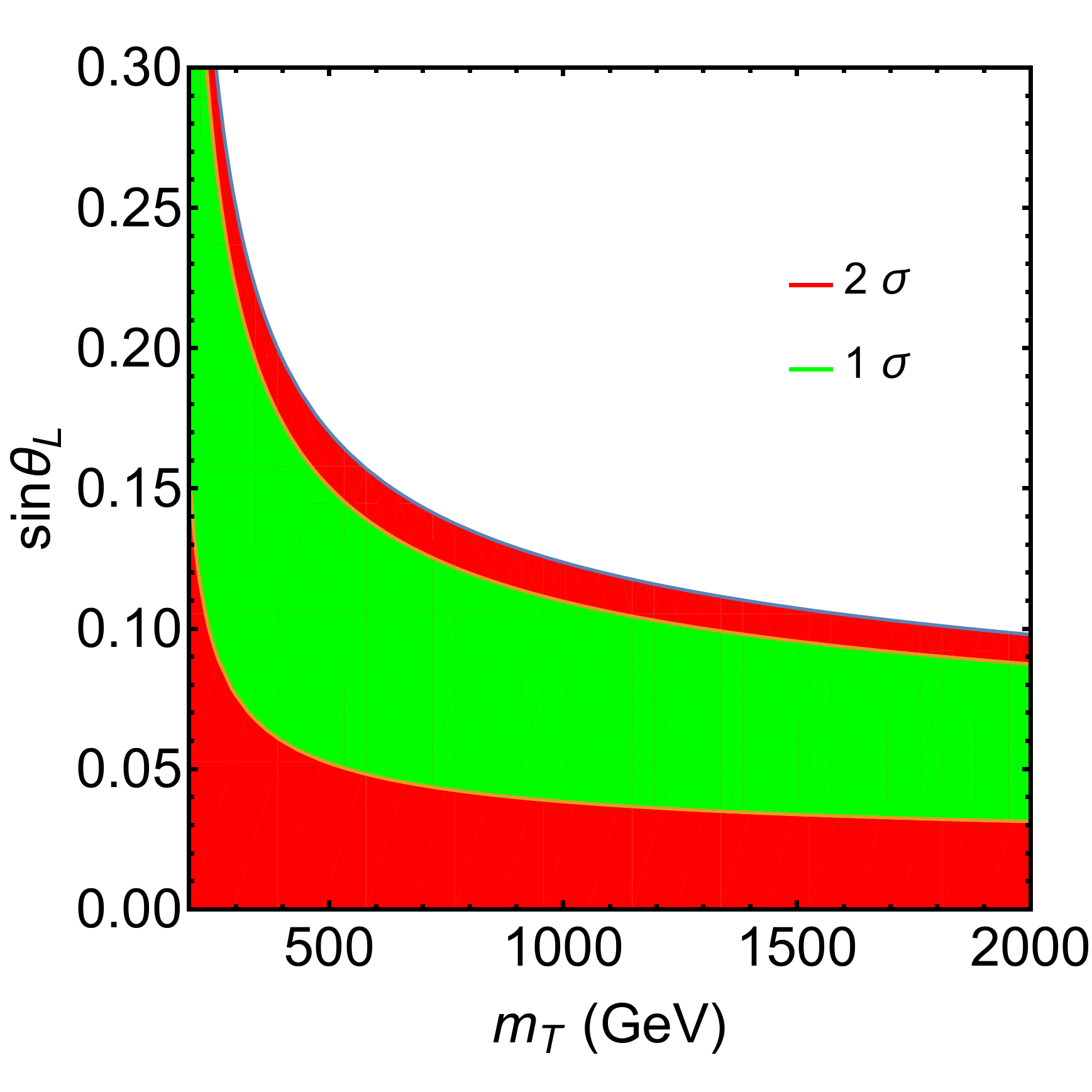}
\caption{The constraints on $m_T,s_L$ from $\chi^2$-fit of the $S,T$ parameters. Here the green and red areas are allowed at $1\sigma,2\sigma$ CL, respectively.}\label{fig:constraints:EWPO}
\end{center}
\end{figure}
In this paper, we choose the parameters to be $m_Z=91.1876\mathrm{GeV},m_W=80.387\mathrm{GeV},m_h=125.09\mathrm{GeV},
m_t=172.74\mathrm{GeV},G_F=1.1664\times10^{-5}\mathrm{GeV}^{-2}$, and $c_W=m_W/m_Z$ \cite{Tanabashi:2018oca}. In Fig.~\ref{fig:constraints:EWPO}, we get the constraints from the global fits of $S,T$ parameters \footnote{$t\textrm{-}T$ mixing will also enter into $Zb\bar{b}$ coupling through one-loop correction, but here we will not consider them anymore.}.
\subsection{Constraints from top physics}
There are also constraints from the $tbW$ anomalous coupling \cite{Cao:2015doa}, which gives the bound $V_{tb}\geq0.92$ at $95\%$ CL assuming $V_{tb}\leq1$ \cite{Khachatryan:2014iya}. Then, we have $s_L\leq\sqrt{1-V_{tb}}\approx0.3$.
\subsection{Constraints from Higgs physics}
In App.~\ref{app:h2gamgam}, we give the exhaustive computations and analyses in both the SM and new physics model. When we take $\kappa_t=c_L^2$ and $y_T=-\frac{m_T}{v}s_L^2$ naively, the following expressions are obtained:
\begin{align}
\mu_{\gamma\gamma}&\equiv\frac{\sigma(gg\rightarrow h)\Gamma(h\rightarrow\gamma\gamma)}{\sigma^{SM}(gg\rightarrow h)\Gamma^{SM}(h\rightarrow\gamma\gamma)}=\frac{\Gamma(h\rightarrow gg)\Gamma(h\rightarrow\gamma\gamma)}{\Gamma^{SM}(h\rightarrow gg)\Gamma^{SM}(h\rightarrow\gamma\gamma)}\nonumber\\
&=|c_L^2+s_L^2\frac{F_f(\tau_T)}{F_f(\tau_t)}|^2~\frac{|N_t^CQ_t^2[c_L^2F_f(\tau_t)+s_L^2F_f(\tau_T)]
+F_W(\tau_W)|^2}{|N_t^CQ_t^2F_f(\tau_t)
+F_W(\tau_W)|^2}.
\end{align}

\begin{figure}[!ht]
\begin{center}
\includegraphics[scale=0.35]{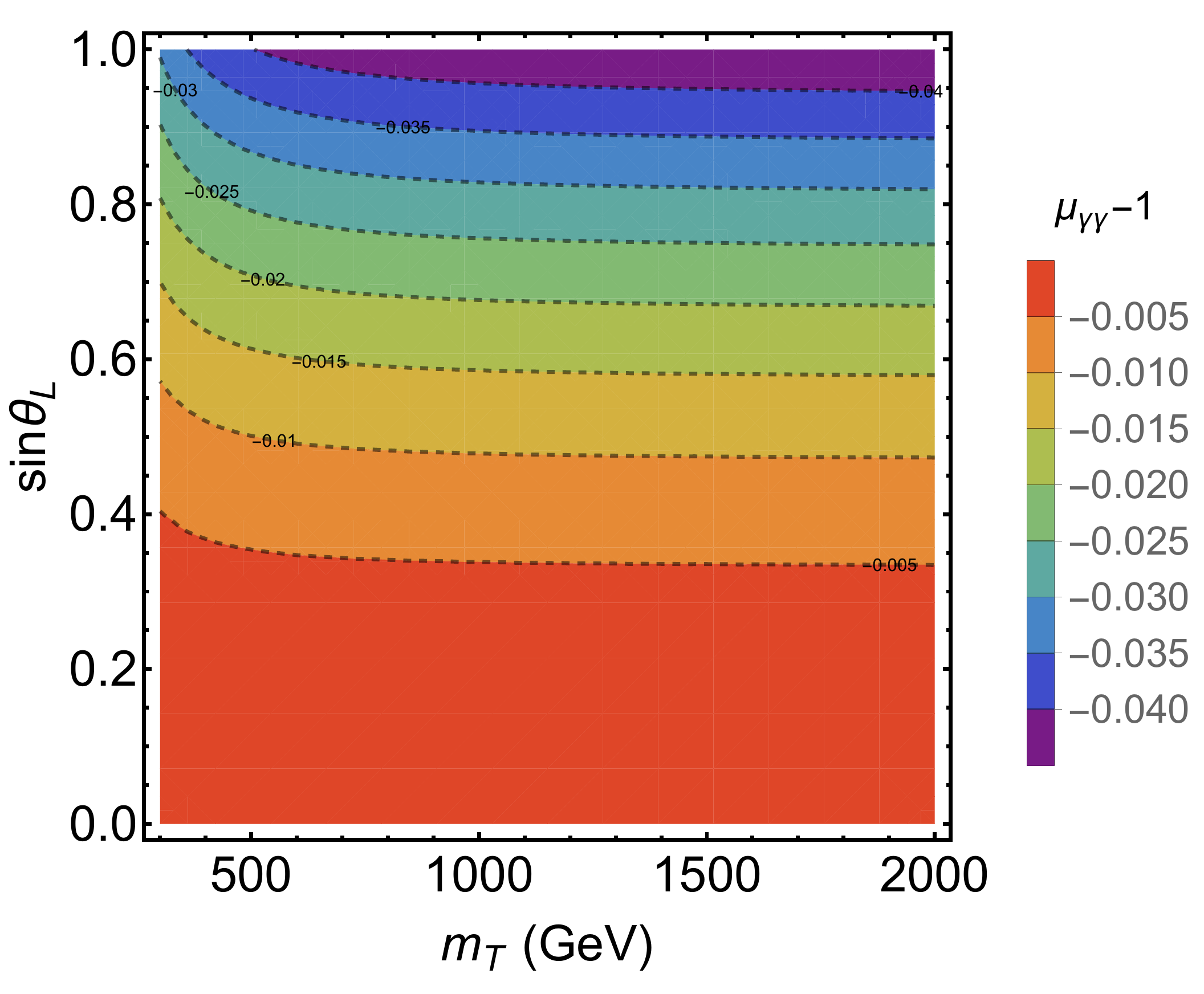}
\caption{The contour plot of the Higgs signal strength deviation for the $gg\rightarrow h\rightarrow\gamma\gamma$ channel in the $m_T-s_L$ plane.}\label{fig:Higgs:SS}
\end{center}
\end{figure}

In Fig.~\ref{fig:Higgs:SS}, we show the contour plot of $(\mu_{\gamma\gamma}-1)$ in the parameter space of $m_T,s_L$. We find that the typical deviation $(\mu_{\gamma\gamma}-1)$ is at the level of $-0.5\%\sim-5\%$, which is within the precision of current measurements \cite{Sirunyan:2018koj, *Aad:2019mbh}. As with the results in Ref.~\cite{Chen:2017hak}, the constraints from Higgs signal strength are quite loose.

\subsection{Constraints from EDM}\label{subsec:EDM}
If there exist $CP$ violation in the FCN interactions, it will contribute to the electron electric dipole moment (EDM). The neutron EDM and chromo EDM (CEDM) will also be affected. Then, the imaginary parts of $y_{L,R}^{tT}$ can be constrained. Here, $CP$ violation is only from the FCNY interactions.

\begin{figure}[!h]
\begin{center}
\includegraphics[scale=0.35]{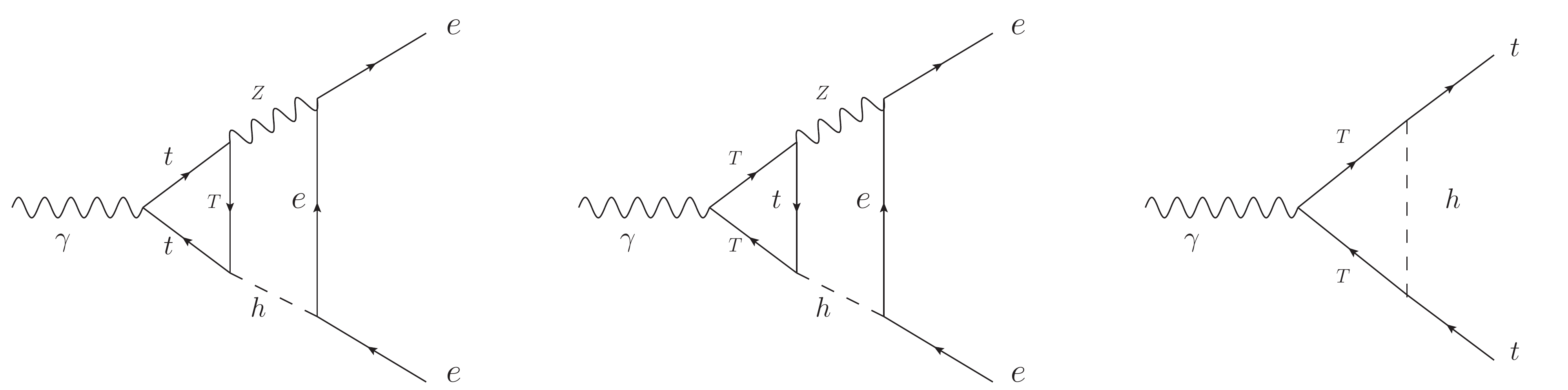}
\caption{The Barr-Zee diagrams contributing to the electron EDM (left, middle) and the Feynman diagrams contributing to the top quark EDM (right). For the fermion loops, counter-clockwise diagrams should be included.}\label{fig:EDM}
\end{center}
\end{figure}
Firstly, the FCN couplings can alter the electron EDM through Barr-Zee diagrams at two-loop level \cite{Barr:1990vd} (see the left and middle diagrams of Fig.~\ref{fig:EDM}). Here, the contributions originate from the $Z$ boson, because there are no FCN couplings for the photon. Due to the $C$ invariance, only vectorial couplings can contribute \cite{Barr:1990vd}. Now, we can make a sketchy estimation. Compared to the photon diagram, $Z$ boson mediated Barr-Zee diagrams are suppressed by 
$\lambda^2\equiv\frac{g_L^{tT}+g_R^{tT}}{2Q_t}\frac{1-4s_W^2}{4s_Wc_W}\sim0.01$. The $CP$ violated $htt$ coupling has been bounded as $|\widetilde{\kappa}_t|<0.01$ \cite{Brod:2013cka}, which comes from the ACME experiment with the electron EDM limit $|d_e|<8.7\times10^{-29}e\cdot$cm \cite{Baron:2013eja}. Currently, the limit is improved to be $|d_e|<1.1\times10^{-29}e\cdot$cm \cite{Andreev:2018ayy}; then, we can rescale the limit of $\widetilde{\kappa}_t$ as $|\widetilde{\kappa}_t|<1.26\times10^{-3}$. From a naive analog, the constraints on FCNY couplings are typically $1.26\times10^{-3}/\lambda^2\sim\mathrm{O}(0.1)$. But this argument is not persuasive, because the two-loop results are unknown for these FCN coupling mediated diagrams. Therefore, we need to resort to other methods. 

Secondly, the FCN couplings can be constrained from top quark EDM and CEDM. The top quark EDM is constrained to be $|d_t^{EDM}|<5\times10^{-20}e\cdot$cm at $90\%$ CL \cite{Hewett:1993em, CorderoCid:2007uc, Kamenik:2011dk, Cirigliano:2016njn, Cirigliano:2016nyn} with the ACME results \cite{Baron:2013eja}. Similarly, we can rescale the limit of the top quark EDM to be $|d_t^{EDM}|<6.3\times10^{-21}e\cdot$cm or $|m_td_t^{EDM}/e|<5.5\times10^{-5}$ with the improved data \cite{Andreev:2018ayy}. The severe constraint on top quark CEDM is inferred from the neutron EDM with the magnitude of $|d_t^{CEDM}|<2.1\times10^{-19}$cm or $|m_td_t^{CEDM}|<1.9\times10^{-3}$ at $90\%$ CL \cite{Kamenik:2011dk, Baker:2006ts, Afach:2015sja, Chien:2015xha, Cirigliano:2016njn, Cirigliano:2016nyn}. In the right diagram of Fig.~\ref{fig:EDM}, we show the Feynman diagram contributing to the top quark EDM. When the photon is replaced by a gluon, we can get the contribution to top CEDM. The interactions induced at one-loop have the following form:
\begin{align}
\mathcal{L}\supset-\frac{i}{2}d_t^{EDM}\bar{t}\sigma^{\mu\nu}\gamma^5tF_{\mu\nu}-\frac{ig_s}{2}d_t^{CEDM}\bar{t}\sigma^{\mu\nu}t^a\gamma^5tG_{\mu\nu}^a.
\end{align}
The expressions of $d_t^{EDM}$ and $d_t^{CEDM}$ are computed as
\begin{align}
&d_t^{EDM}=\frac{eQ_Tm_T[y_R^{tT}(y_L^{tT})^*-y_L^{tT}(y_R^{tT})^*]}{16\pi^2}C_1,\quad d_t^{CEDM}=\frac{m_T[y_R^{tT}(y_L^{tT})^*-y_L^{tT}(y_R^{tT})^*]}{16\pi^2}C_1,
\end{align}
where $C_1$ is defined as
\begin{align*}
&C_1=\frac{1}{4m_t^2}[B_0(m_t^2,m_T^2,m_h^2)-B_0(0,m_T^2,m_T^2)+(m_T^2-m_t^2-m_h^2)C_0(m_t^2,0,m_t^2,m_h^2,m_T^2,m_T^2)].
\end{align*}
$[y_R^{tT}(y_L^{tT})^*-y_L^{tT}(y_R^{tT})^*]$ can also be rewritten as $2i(\mathrm{Re}y_L^{tT}\mathrm{Im}y_R^{tT}-\mathrm{Re}y_R^{tT}\mathrm{Im}y_L^{tT})$; thus, $d_t^{EDM},~d_t^{CEDM}$ will vanish if the imaginary parts of $y_{L,R}^{tT}$ are both turned off. If we take $m_T=700~\mathrm{GeV}$, top EDM and CEDM set the upper limits of $|y_R^{tT}(y_L^{tT})^*-y_L^{tT}(y_R^{tT})^*|$ to be 0.21 and 4.9, respectively. If we take $m_T=400~\mathrm{GeV}$, the corresponding upper limits of $|y_R^{tT}(y_L^{tT})^*-y_L^{tT}(y_R^{tT})^*|$ are 0.12 and 2.8, respectively. Thus, top quark EDM will give much stronger constraints than top CEDM.

\section{Partial decay width formula of $h\rightarrow\gamma Z$}\label{sec:width}
\subsection{SM result}
\begin{figure}[!h]
\includegraphics[scale=0.45]{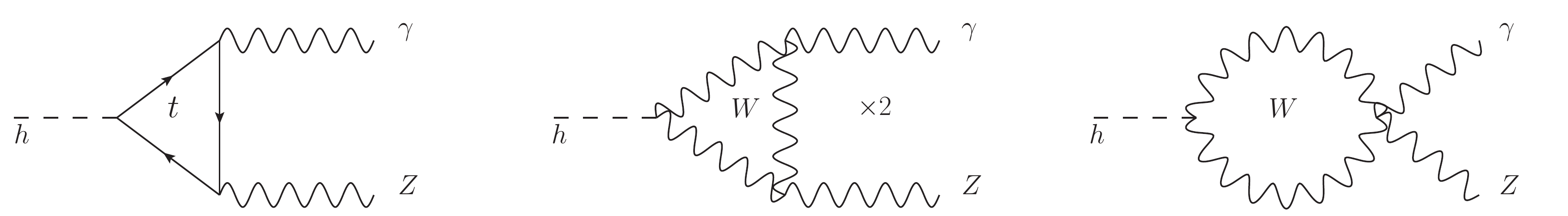}
\caption{Typical Feynman diagrams contributing to $h\rightarrow\gamma Z$ decay in the SM. For the fermion loops, counter-clockwise diagrams should be included.}\label{fig:Feyn:SM}
\end{figure}
There are contributions from top and $W$ loops for the $h\rightarrow\gamma Z$ decay. In Fig.~\ref{fig:Feyn:SM}, we show the Feynman diagrams drawn by JaxoDraw \cite{Binosi:2008ig}. The partial decay width in SM is computed as \cite{Bergstrom:1985hp, Gunion:1989we, Djouadi:1996yq, Djouadi:2005gi, Boradjiev:2017khm}
\begin{align}
&\Gamma^{SM}(h\rightarrow\gamma Z)=\frac{G_F\alpha^2m_h^3}{64\sqrt{2}\pi^3}(1-\frac{m_Z^2}{m_h^2})^3|\sum_f(2N_f^CQ_f)\frac{I_3^f-2Q_fs_W^2}{s_Wc_W}A_f(\tau_f,\lambda_f)+A_W(\tau_W,\lambda_W)|^2.
\end{align}
Here, $\tau_i$ and $\lambda_i$ are defined as
\begin{align}
\tau_f=\frac{4m_f^2}{m_h^2},\tau_W=\frac{4m_W^2}{m_h^2},\lambda_f=\frac{4m_f^2}{m_Z^2},\lambda_W=\frac{4m_W^2}{m_Z^2}.
\end{align}
and the $A_f,A_W$ are defined as
\begin{align}
&A_f(\tau_f,\lambda_f)\equiv I_1(\tau_f,\lambda_f)-I_2(\tau_f,\lambda_f),\nonumber\\
&A_W(\tau_W,\lambda_W)\equiv\frac{1}{t_W}\{[(1+\frac{2}{\tau_W})t_W^2-(5+\frac{2}{\tau_W})]I_1(\tau_W,\lambda_W)+4(3-t_W^2)I_2(\tau_W,\lambda_W)\},\nonumber\\
&I_1(\tau,\lambda)=\frac{\tau\lambda}{2(\tau-\lambda)}+\frac{\tau^2\lambda^2}{2(\tau-\lambda)^2}[f(\tau)-f(\lambda)]+\frac{\tau^2\lambda}{(\tau-\lambda)^2}[g(\tau)-g(\lambda)],\nonumber\\
&I_2(\tau,\lambda)=-\frac{\tau\lambda}{2(\tau-\lambda)}[f(\tau)-f(\lambda)].
\end{align}
Here, $f(\tau)$ is given in App.~\ref{app:h2gamgam} and $g(\tau)$ is defined as
\begin{align}
&g(\tau)\equiv
              \begin{cases}
              \sqrt{\tau-1}\arcsin(\frac{1}{\sqrt{\tau}}), &\mathrm{for}~\tau\geq1\\
              \frac{1}{2}\sqrt{1-\tau}[\log\frac{1+\sqrt{1-\tau}}{1-\sqrt{1-\tau}}-i\pi], &\mathrm{for}~\tau<1
              \end{cases}.
\end{align}
The fermionic part is dominated by the top quark because of the largest Yukawa coupling. Numerically, we can get $(2N_t^CQ_t)\frac{I_3^t-2Q_ts_W^2}{s_Wc_W}A_f(\tau_t,\lambda_t)\sim-0.65,A_W(\tau_W,\lambda_W)\sim12.03$, which means the gauge boson contributions are almost 18.5 times larger than the fermionic ones. It is obvious that the fermionic part and gauge boson part interfere destructively in the SM.
\subsection{New physics result}\label{subsec:h2gamZ:NP}
\begin{figure}[!h]
\includegraphics[scale=0.45]{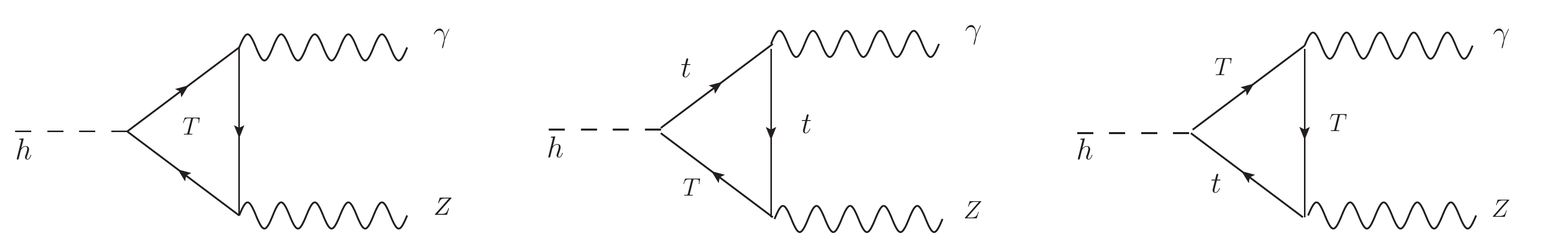}
\caption{Possible new fermion contributions to the $h\rightarrow\gamma Z$ decay. For the fermion loops, counter-clockwise diagrams should be included.}\label{fig:Feyn:BSM}
\end{figure}

$h\rightarrow\gamma Z$ decay has already been considered in many models, for example, composite Higgs models \cite{Azatov:2013ura, Cao:2018cms}, minimal supersymmetric standard model (MSSM) \cite{Djouadi:2005gj, Cao:2013ur}, next-to-MSSM (NMSSM) \cite{Cao:2013ur, Belanger:2014roa}, extended scalar sector models \cite{Chiang:2012qz, Swiezewska:2012eh, Chen:2013vi}, and other new physics models \cite{Chen:2013dh}. In VLQ models, there are additional fermion contributions: pure new quark loops, loops with both SM and new quarks (see Fig.~\ref{fig:Feyn:BSM}). The latter will be induced by the FCN interactions. Such off-diagonal contributions are always ignored in most studies \cite{Djouadi:1996yq, Azatov:2013ura}, because they are small compared to the diagonal terms. As a second thought, this channel can be sensitive to large non-diagonal couplings. Here, we do not enumerate models with more fermions, where the effects of non-diagonal couplings will be diluted or concealed. Besides, we only focus on the cases in which the scalar sector is extended with real gauge singlet scalars. In more complex scalar sector models, the charged Higgs contributions will also attenuate the flavour off-diagonal contributions.

Now, let us consider the partial decay width of $h\rightarrow\gamma Z$ with the general interactions in Eq.~\eqref{eqn:frame:simplify}. Due to $U_{EM}(1)$ gauge symmetry, the $h\rightarrow\gamma Z$ amplitude possesses the following tensor structure \footnote{During the calculations, we have used the FeynCalc to simplify the results \cite{Mertig:1990an, Shtabovenko:2016sxi}.}:
\begin{align}
&i\mathcal{M}=i\epsilon_{\mu}(p_1)\epsilon_{\nu}(p_2)[(p_2^{\mu}p_1^{\nu}-p_1\cdot p_2g^{\mu\nu})\mathcal{A}+\epsilon^{\mu\nu p_1p_2}B]~(\epsilon^{\mu\nu p_1p_2}\equiv\epsilon^{\mu\nu\rho\sigma}p_{1,\rho}p_{2,\sigma}),\nonumber\\
&\mathcal{A}\equiv\frac{e^2}{8\pi^2v}(\mathcal{A}_W+\mathcal{A}_t+\mathcal{A}_T+\mathcal{A}_{tT}),\qquad \mathcal{B}\equiv\frac{e^2}{8\pi^2v}\mathcal{B}_{tT},
\end{align}
where $\mathcal{A}_W,\mathcal{A}_t,\mathcal{A}_T,\mathcal{A}_{tT}$ denote the contributions from $W$ boson, top quark, $T$ quark and $t-T$ mixed loops, respectively. Their expressions are given as
\begin{align*}
&\mathcal{A}_W=A_W(\tau_W,\lambda_W),\nonumber\\
&\mathcal{A}_t=2N_t^CQ_t(g_{L}^{t}+g_{R}^{t})\kappa_tA_f(\tau_t,\lambda_t)=2N_t^CQ_t\kappa_t\frac{\frac{1}{2}c_L^2-\frac{4}{3}s_W^2}{s_Wc_W}A_f(\tau_t,\lambda_t),\nonumber\\
&\mathcal{A}_T=-2N_T^CQ_T\frac{y_{T}v}{m_T}(g_{L}^{T}+g_{R}^{T})A_f(\tau_T,\lambda_T)=-2N_T^CQ_T\frac{y_{T}v}{m_T}\frac{\frac{1}{2}s_L^2-\frac{4}{3}s_W^2}{s_Wc_W}A_f(\tau_T,\lambda_T),
\end{align*}
\begin{align}
&\mathcal{A}_{tT}=-4N_T^CQ_Tg_{L}^{tT}\frac{v}{m_h^2-m_Z^2}\{m_t\mathrm{Re}(y_L^{tT})[(\frac{m_h^2-m_Z^2}{2}-m_t^2)C_0(0,m_Z^2,m_h^2,m_t^2,m_t^2,m_T^2)\nonumber\\
&-m_T^2C_0(0,m_Z^2,m_h^2,m_T^2,m_T^2,m_t^2)-m_Z^2\frac{B_0(m_h^2,m_t^2,m_T^2)-B_0(m_Z^2,m_t^2,m_T^2)}{m_h^2-m_Z^2}-1]\nonumber\\
&+m_T\mathrm{Re}(y_R^{tT})[(\frac{m_h^2-m_Z^2}{2}-m_T^2)C_0(0,m_Z^2,m_h^2,m_T^2,m_T^2,m_t^2)\nonumber\\
&-m_t^2C_0(0,m_Z^2,m_h^2,m_t^2,m_t^2,m_T^2)-m_Z^2\frac{B_0(m_h^2,m_t^2,m_T^2)-B_0(m_Z^2,m_t^2,m_T^2)}{m_h^2-m_Z^2}-1]\}.
\end{align}
Similarly, the expression of $\mathcal{B}_{tT}$ is given as
\begin{align}
&\mathcal{B}_{tT}=-2N_T^CQ_Tg_{L}^{tT}v[m_t\mathrm{Im}(y_L^{tT})C_0(0,m_Z^2,m_h^2,m_t^2,m_t^2,m_T^2)-m_T\mathrm{Im}(y_R^{tT})C_0(0,m_Z^2,m_h^2,m_T^2,m_T^2,m_t^2)].
\end{align}

Taking the mass of $t,T$ quarks to be infinity, $\mathcal{A}_t,~\mathcal{A}_T$ can be expanded as
\begin{align}
&\mathcal{A}_t\approx-\frac{2}{3}N_t^CQ_t\kappa_t\frac{\frac{1}{2}c_L^2-\frac{4}{3}s_W^2}{s_Wc_W}[1+\frac{7m_h^2+11m_Z^2}{120m_t^2}+\mathcal{O}(\frac{m_h^4,m_h^2m_Z^2,m_Z^4}{m_t^4})],\nonumber\\
&\mathcal{A}_T\approx-\frac{2}{3}N_T^CQ_T(-\frac{y_{T}v}{m_T})\frac{\frac{1}{2}s_L^2-\frac{4}{3}s_W^2}{s_Wc_W}[1+\frac{7m_h^2+11m_Z^2}{120m_T^2}+\mathcal{O}(\frac{m_h^4,m_h^2m_Z^2,m_Z^4}{m_T^4})],\nonumber\\
&\mathcal{A}_t+\mathcal{A}_T\approx-\frac{2N_T^CQ_T}{3s_Wc_W}\left[\kappa_t(\frac{1}{2}c_L^2-\frac{4}{3}s_W^2)(1+\frac{7m_h^2+11m_Z^2}{120m_t^2})+(-\frac{y_{T}v}{m_T})(\frac{1}{2}s_L^2-\frac{4}{3}s_W^2)(1+\frac{7m_h^2+11m_Z^2}{120m_T^2})\right].
\end{align}
For the $\frac{1}{m_{t,T}^2}$ suppressed contributions, we can get $\frac{7m_h^2+11m_Z^2}{120m_t^2}\approx5.5\%,\frac{7m_h^2+11m_Z^2}{120m_T^2}\lesssim1\%$ if $m_T\gtrsim400\mathrm{GeV}$. The expansion of $A_{tT}$ is a little bit complicated:
\begin{align}
&\mathcal{A}_{tT}\approx-4N_T^CQ_Tg_{L}^{tT}\frac{v}{m_h^2-m_Z^2}\{m_t\mathrm{Re}(y_L^{tT})[\frac{m_h^2-m_Z^2}{m_T^2}\frac{(1-r_{tT}^2)(3-r_{tT}^2)+2\log r_{tT}^2}{4(1-r_{tT}^2)^3}+\mathcal{O}(\frac{m_h^4,m_h^2m_Z^2,m_Z^4}{m_t^4,m_t^2m_T^2,m_T^4})]\nonumber\\
	&+m_T\mathrm{Re}(y_R^{tT})[\frac{m_h^2-m_Z^2}{m_T^2}\frac{2r_{tT}^4\log r_{tT}^2-(1-r_{tT}^2)(1-3r_{tT}^2)}{4(1-r_{tT}^2)^3}+\mathcal{O}(\frac{m_h^4,m_h^2m_Z^2,m_Z^4}{m_t^4,m_t^2m_T^2,m_T^4})]\}\nonumber\\
&\approx-N_T^CQ_Tg_{L}^{tT}\frac{v}{m_T^2}[m_t\mathrm{Re}(y_L^{tT})(3+2\log r_{tT}^2)-m_T\mathrm{Re}(y_R^{tT})].
\end{align}
Similarly, we can expand $B_{tT}$ as
\begin{align}
&\mathcal{B}_{tT}\approx-2N_T^CQ_Tg_{L}^{tT}\frac{v}{m_T^2}\{m_t\mathrm{Im}(y_L^{tT})[1+\log r_{tT}^2+\frac{m_h^2+m_Z^2}{m_T^2}\frac{5+2(1+2r_{tT}^2)\log r_{tT}^2}{4}]\nonumber\\
&+m_T\mathrm{Im}(y_R^{tT})[1+r_{tT}^2\log r_{tT}^2+\frac{m_h^2+m_Z^2}{m_T^2}\frac{1+4r_{tT}^2\log r_{tT}^2}{4}]\}.
\end{align}
In Tab.~\ref{tab:Af}, we list the expressions of $A_t+A_T,A_{tT},B_{tT}$ in three models, where we have neglected the $\frac{1}{m_{t,T}^2}$ suppressed terms but keep the $\log{r_{tT}^2}$ enhanced terms.
\begin{sloppypar}
\begin{table}[!h]
\begin{tabular}{c|c|c|c}
\hline
\diagbox{\qquad}{\qquad}& $\bar{A}_t+\bar{A}_T$ & $\bar{A}_{tT}$ & $\bar{B}_{tT}$\\
\hline
SM & $1-\frac{8}{3}s_W^2$ & $\times$ & $\times$\\
\hline
VLQT & $1-\frac{8}{3}s_W^2-2s_L^2c_L^2$ & $\frac{3}{2}s_L^2c_L^2[1-r_{tT}^2(3+2\log{r_{tT}^2})]$ & 0\\
\hline
VLQT+S & \makecell{$c_{\theta}(1-\frac{8}{3}s_W^2-2s_L^2c_L^2)$\\$+\frac{v\mathrm{Re}(y_T^S)}{m_T}\frac{s_{\theta}c_R}{c_L}(\frac{8}{3}s_W^2-2s_L^2c_L^2)$} & \makecell{$\frac{3}{2}s_Lc_L[s_Lc_Lc_{\theta}(1-r_{tT}^2(3+2\log{r_{tT}^2}))$\\$+\frac{v\mathrm{Re}(y_T^S)}{m_T}s_{\theta}(s_Lc_R-r_{tT}(3+2\log{r_{tT}^2})s_Rc_L)]$} &$-\frac{3v}{m_T}s_L^2c_Lc_Rs_{\theta}\mathrm{Im}(y_T^S)$\\
\hline
\end{tabular}
\caption{The expressions of $\bar{A}_t+\bar{A}_T,\bar{A}_{tT},\bar{B}_{tT}$ in the SM, VLQT, and VLQT+S. Here, we extract the common factor $-\frac{N_T^CQ_T}{3s_Wc_W}$ for convenience, that is, redefinition of $A(B)$ with $-\frac{N_T^CQ_T}{3s_Wc_W}\bar{A}(\bar{B})$. We take $\bar{A}_T=0$ naively in SM because of the absence of a $T$ quark.}\label{tab:Af}
\end{table}
\end{sloppypar}

The partial decay width formula is computed as
\begin{align}
&\Gamma(h\rightarrow\gamma Z)=\frac{G_F\alpha^2m_h^3}{64\sqrt{2}\pi^3}(1-\frac{m_Z^2}{m_h^2})^3[|A_t+A_T+A_{tT}+A_W(\tau_W,\lambda_W)|^2+|\mathcal{B}_{tT}|^2].
\end{align}
\subsection{Comments from the viewpoint of low energy theorem}
As a matter of fact, we can also understand some behaviors of the $h\rightarrow\gamma Z$ amplitude resorting to the low energy theorem \cite{Ellis:1975ap, Shifman:1979eb}.
Just as the calculation of $h\rightarrow\gamma\gamma$ amplitude from photon self-energy contribution \cite{Carena:2012xa, Cao:2015scs}, we may get the $h\rightarrow\gamma Z$ amplitude through $\gamma-Z$ mixed self-energy contribution \cite{Kniehl:1995tn}. But what confuses us is that there seems no off-diagonal fermion contributions to the $\gamma-Z$ two-point function because photon can only couple to the same flavour particle. The reason is that off-diagonal couplings are proportional to the mixing angle, which is suppressed by the heavy fermion mass. Thus, off-diagonal contributions to the $h\rightarrow\gamma Z$ amplitude vanish in the limit of $p_h\rightarrow0$, consistent with the corollary of low energy theorem. In other words, this channel will give looser constraints on off-diagonal couplings once one flavour of the loop particles becomes heavier.
\section{Numerical results and constraint prospects}\label{sec:numerical}
Just similar to the VLQT model, we take $\kappa_t=c_L^2,y_T=-\frac{m_T}{v}s_L^2$ for simplicity, but let $\mathrm{Re}(y_L^{tT}),\mathrm{Re}(y_R^{tT}),\mathrm{Im}(y_L^{tT}),\mathrm{Im}(y_R^{tT})$ be free. Then, we can choose several benchmark scenarios and estimate the constraints on the magnitude and sign of the FCNY couplings.

Since the branching ratio of $h\rightarrow\gamma Z$ is about $1.5\times10^{-3}$, the modification of $h\gamma Z$ partial decay width will cause negligible effects on the Higgs total width. At the high luminosity LHC (HL-LHC), $h\gamma Z$ coupling can be measured accurately \cite{Cepeda:2019klc, Goertz:2019uek}. The expected $1\sigma$ uncertainty of $Br(h\rightarrow\gamma Z)$ is $19.1\%$ \cite{Cepeda:2019klc}, which gives the following constraint:
\begin{align}\label{eqn:numerical:HLLHC}
|\Gamma(h\rightarrow\gamma Z)/\Gamma^{SM}(h\rightarrow\gamma Z)-1|\leq19.1\%.
\end{align}
It means $|(|A|^2+|B|^2)/|A^{SM}|^2-1|\leq19.1\%$. From now on, we will choose $m_T=400$~GeV and $s_L=0.2$. As mentioned above, there are four interesting parameters: $\mathrm{Re}(y_L^{tT}),\mathrm{Re}(y_R^{tT}),\mathrm{Im}(y_L^{tT}),\mathrm{Im}(y_R^{tT})$. In the following, we will plot the reached two-dimensional parameter space by setting two of them to be zeros or imposing two conditions.

\begin{figure}[!h]
\includegraphics[scale=0.28]{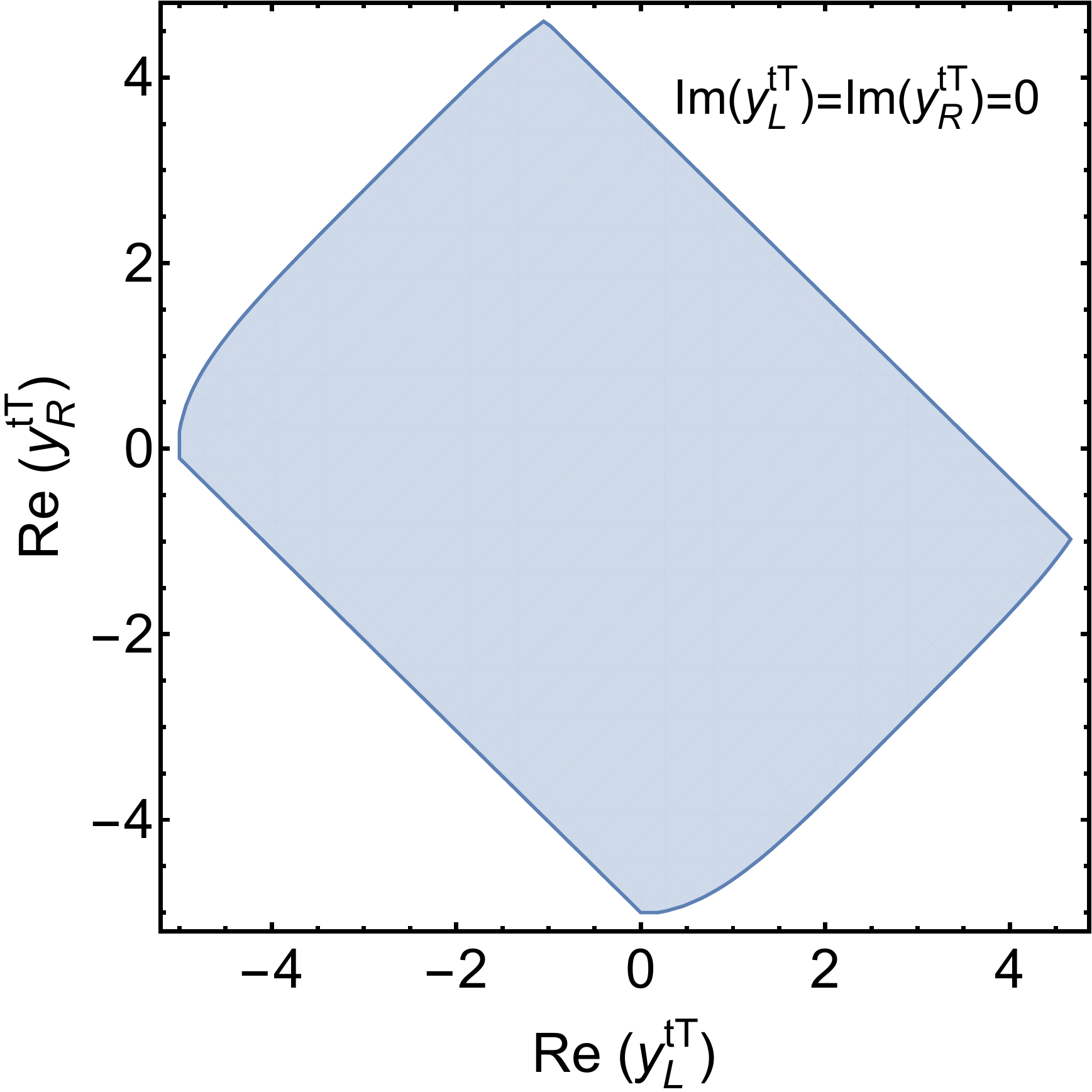}
\includegraphics[scale=0.28]{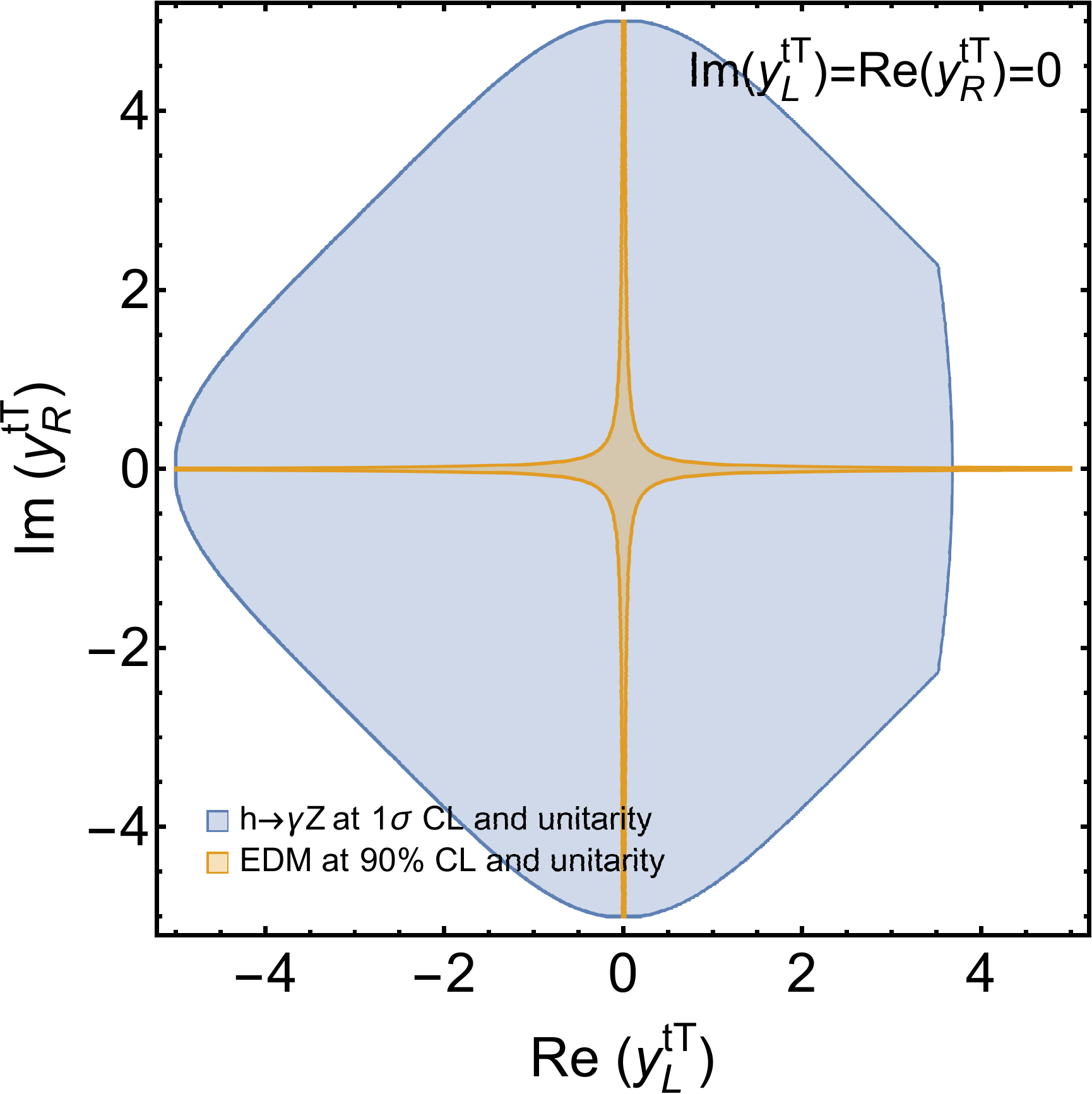}
\includegraphics[scale=0.28]{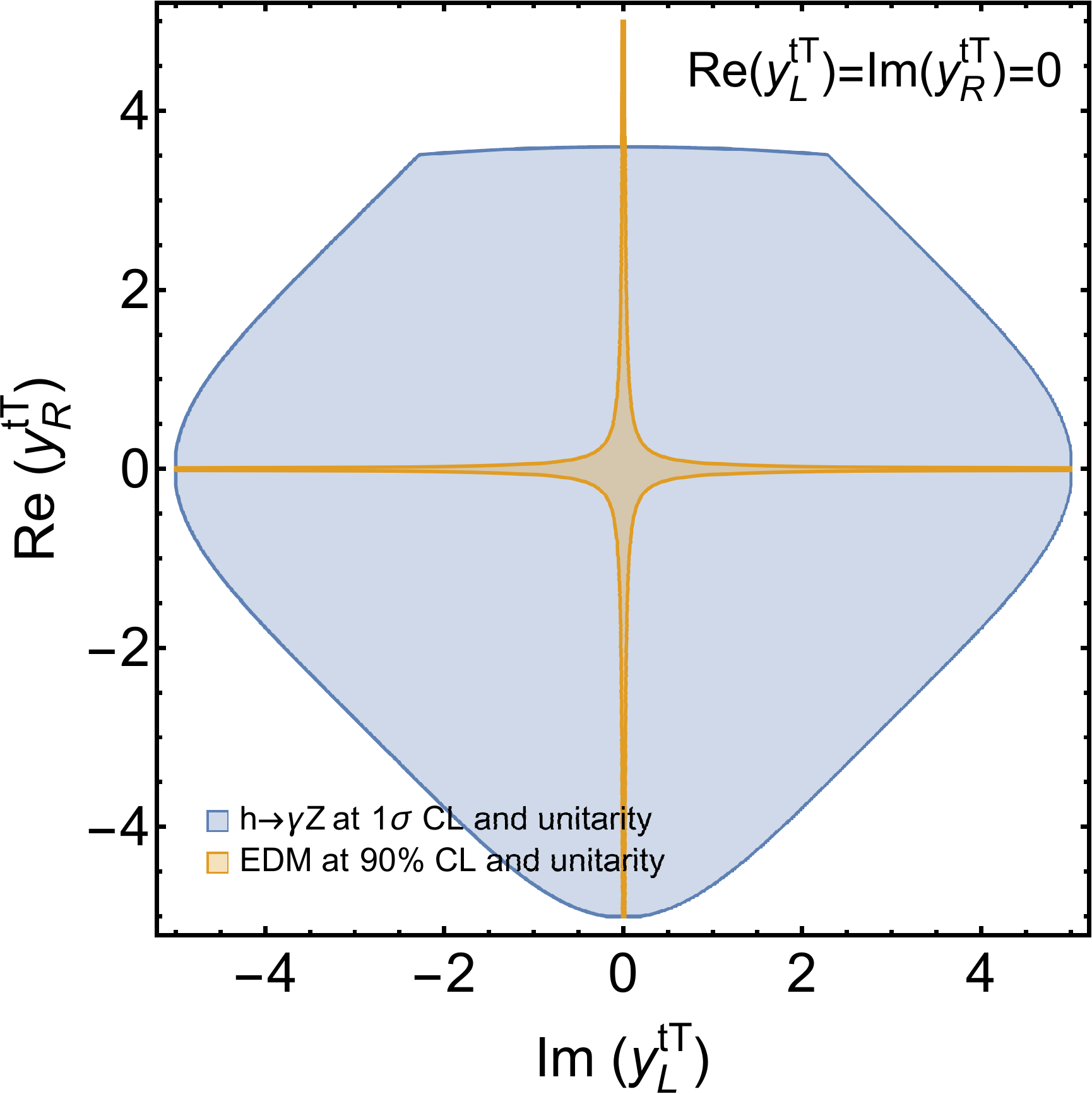}\\
\includegraphics[scale=0.28]{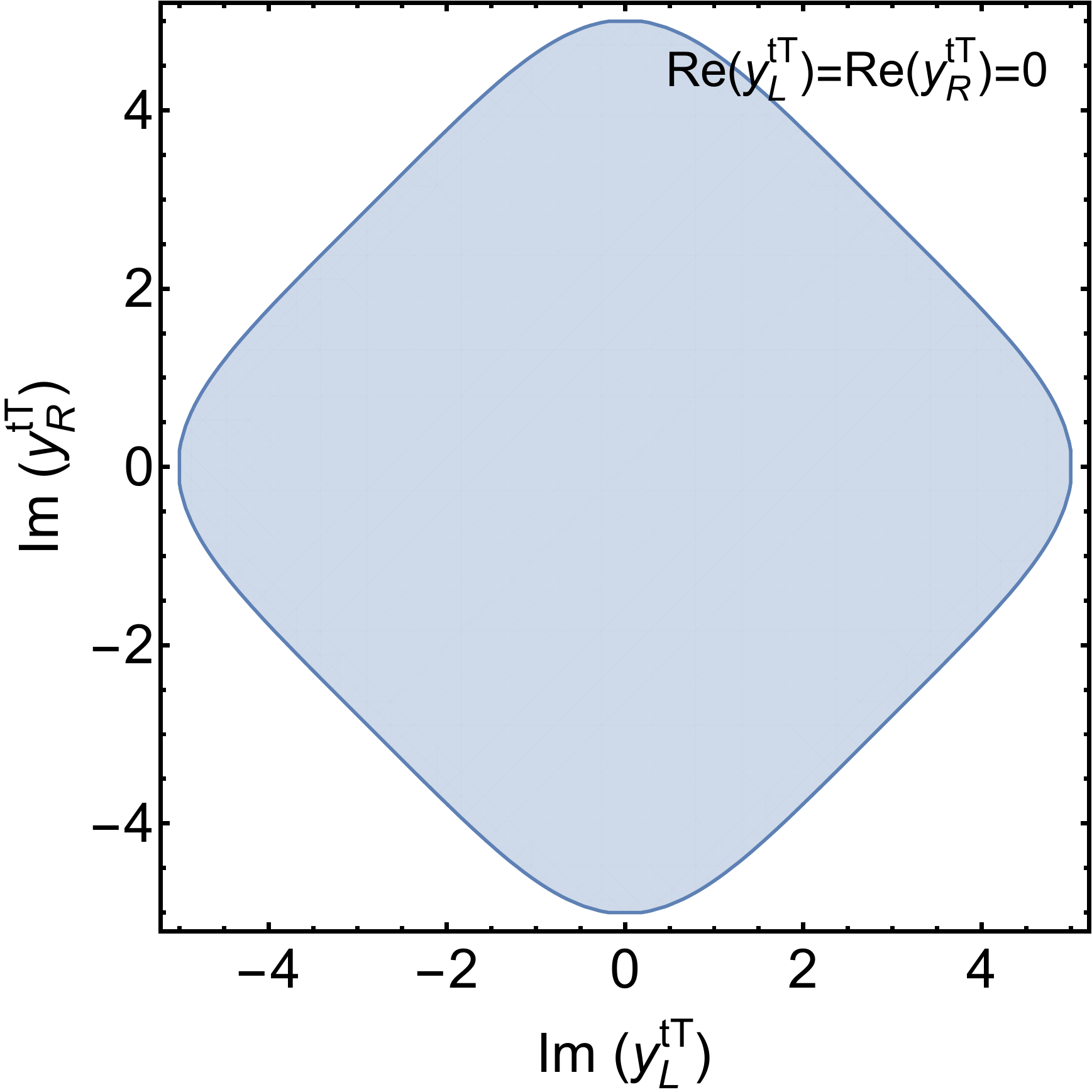}
\includegraphics[scale=0.28]{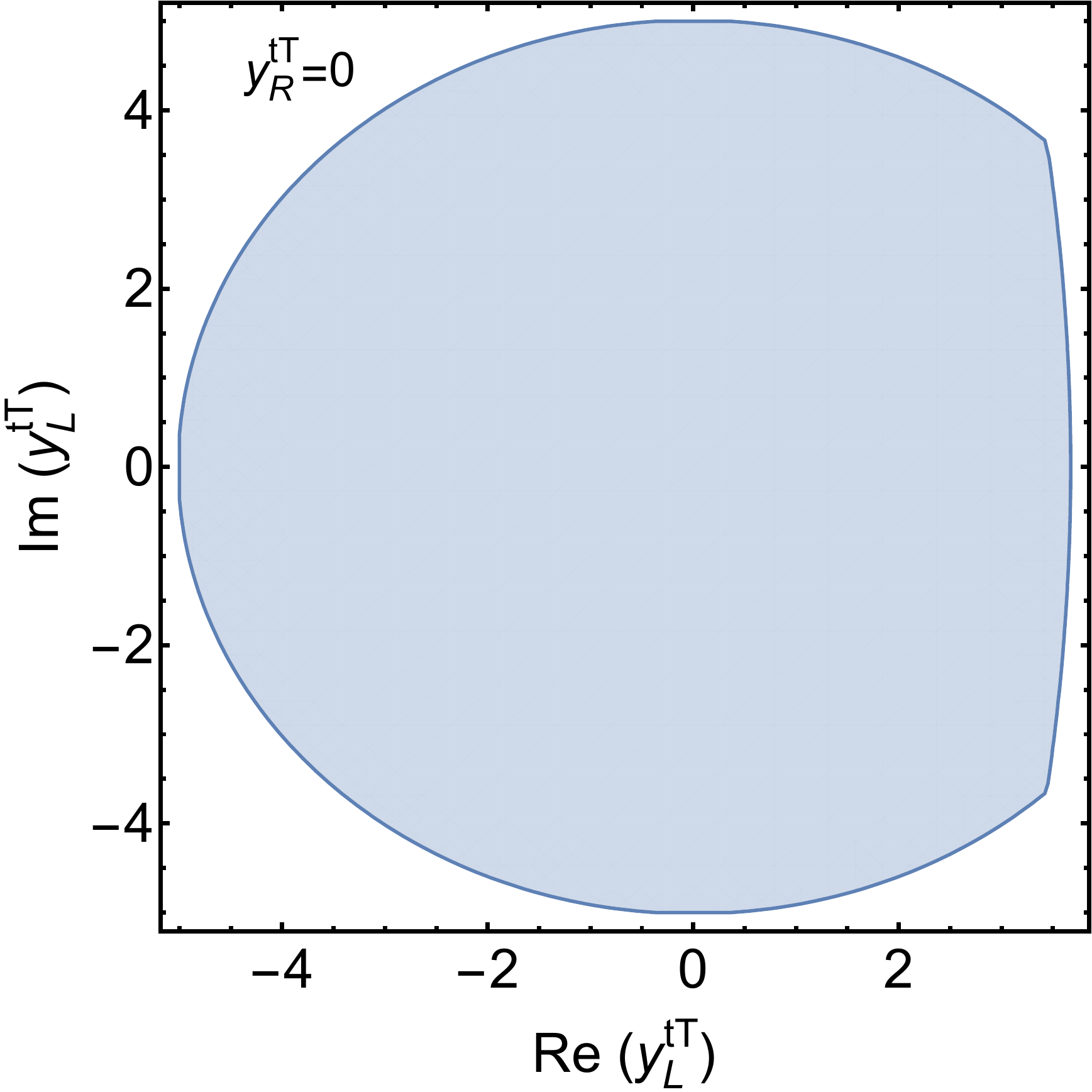}
\includegraphics[scale=0.28]{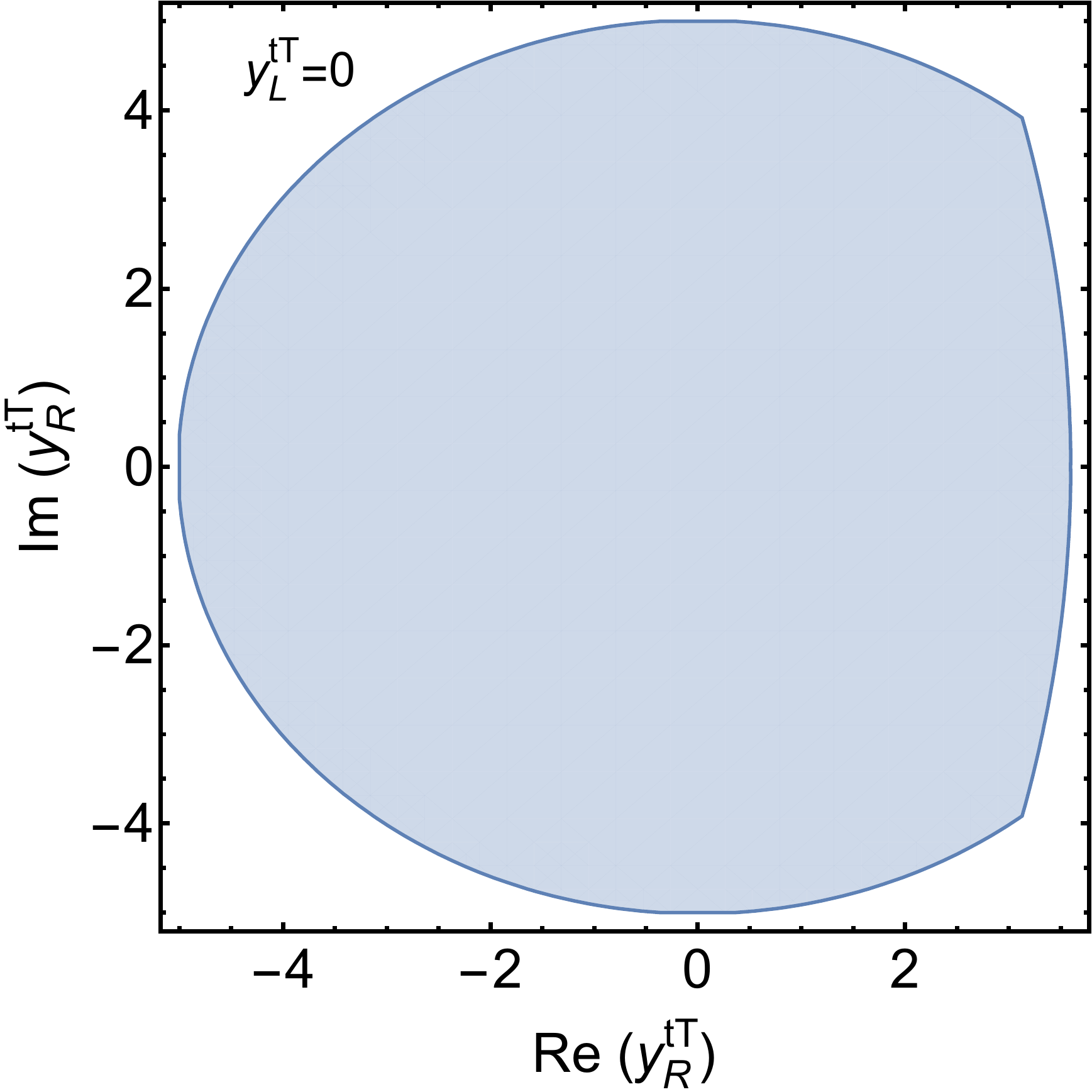}\\
\includegraphics[scale=0.29]{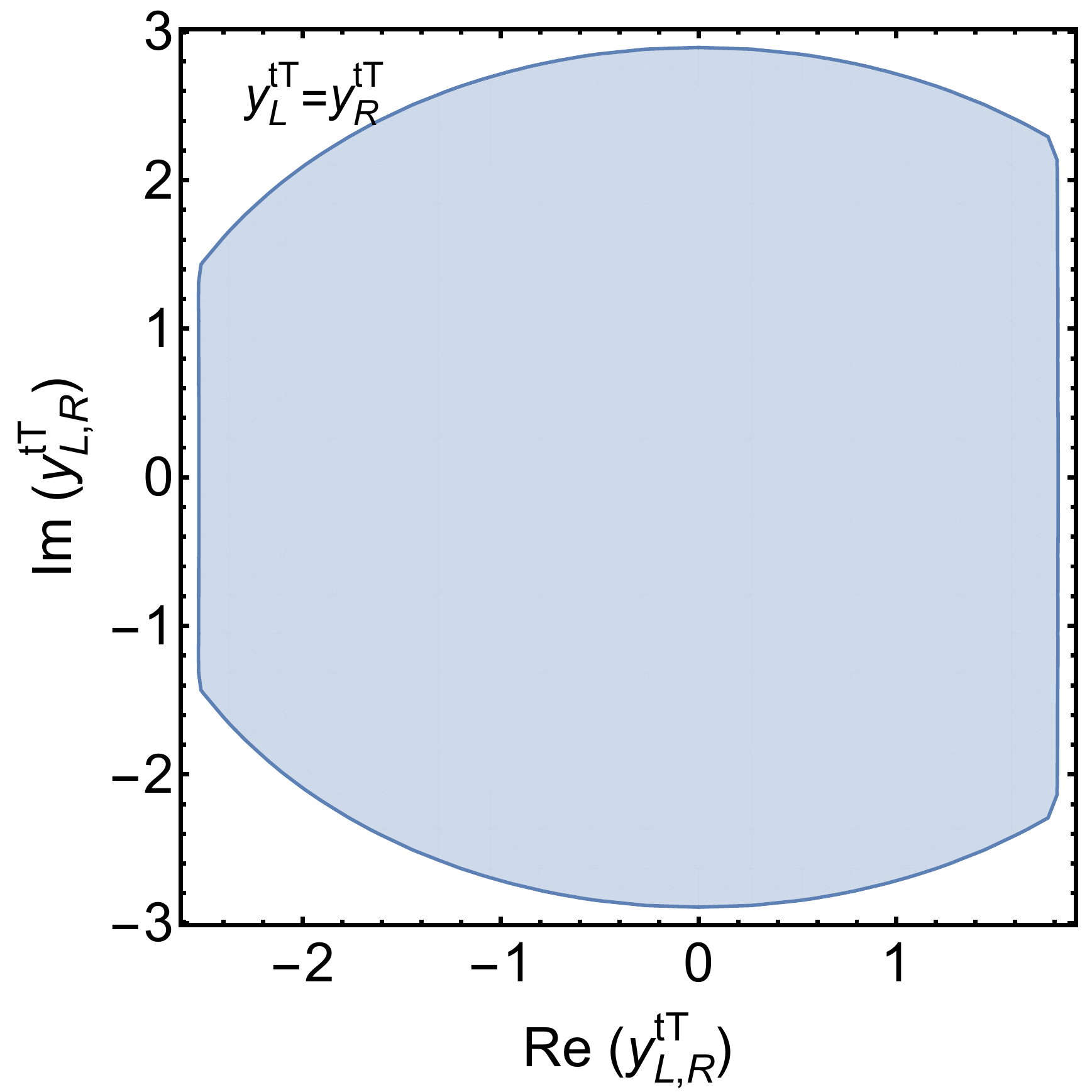}
\includegraphics[scale=0.29]{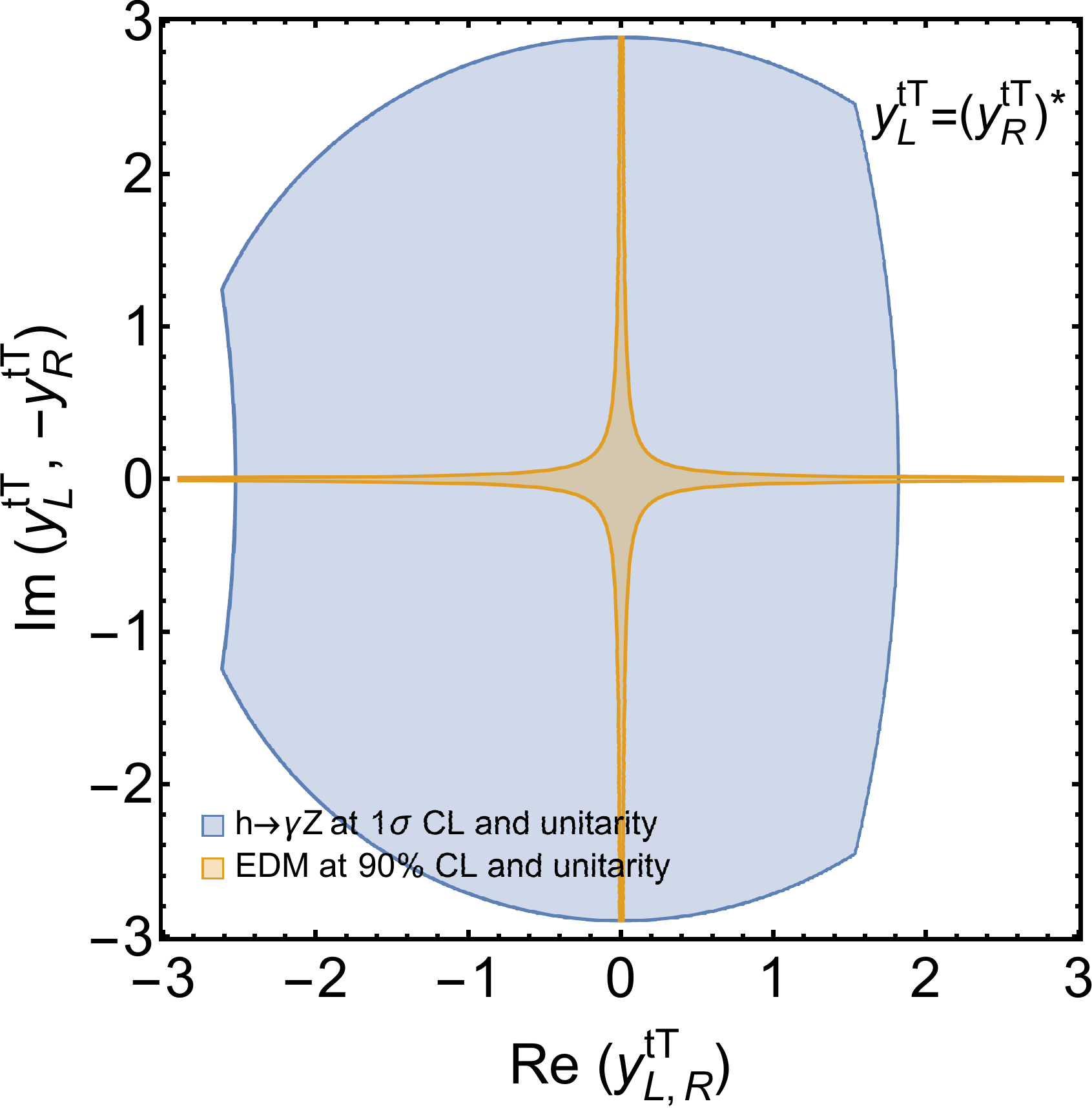}
\caption{The reach regions of $y_L^{tT},y_R^{tT}$ in several scenarios. In the above plots, we take $\mathrm{Im}(y_L^{tT})=\mathrm{Im}(y_R^{tT})=0$ (upper left), $\mathrm{Im}(y_L^{tT})=\mathrm{Re}(y_R^{tT})=0$ (upper central), $\mathrm{Re}(y_L^{tT})=\mathrm{Im}(y_R^{tT})=0$ (upper right), $\mathrm{Re}(y_L^{tT})=\mathrm{Re}(y_R^{tT})=0$ (middle left), $y_R^{tT}=0$ (middle central), $y_L^{tT}=0$ (middle right), $y_L^{tT}=y_R^{tT}$ (lower left) and $y_L^{tT}=(y_R^{tT})^*$ (lower right), respectively. Here only the three scenarios $\mathrm{Im}(y_L^{tT})=\mathrm{Re}(y_R^{tT})=0$, $\mathrm{Re}(y_L^{tT})=\mathrm{Im}(y_R^{tT})=0$, and $y_L^{tT}=(y_R^{tT})^*$ can be constrained by the top quark EDM constraints. The blue region means that it can be reached by the $h\rightarrow\gamma Z$ decay at $1\sigma$ CL and allowed by the unitarity bounds, and the yellow region means that it is allowed by the EDM at $90\%$ CL and unitarity bounds.}\label{fig:numerical:ytT}
\end{figure}
In Fig.~\ref{fig:numerical:ytT}, we plot the parameter space regions allowed by perturbative unitarity in Eq.~\eqref{eqn:unitarity} and the expected constraints at HL-LHC in Eq.~\eqref{eqn:numerical:HLLHC} in different scenarios. The reach regions are shown in blue at $1\sigma$ CL, and the $2\sigma$ bounds in $h\rightarrow\gamma Z$ decay are weaker than the unitarity constraints. When evaluating the scalar loop functions, LoopTools is employed \cite{Hahn:1998yk}. In the first plot, we can find that $h\rightarrow\gamma Z$ decay gives a little stronger constraints than perturbative unitarity in the first and third quadrants in the case of vanishing imaginary parts of $y_L^{tT},y_R^{tT}$. In the presence of imaginary part, the real part can be constrained to be less than 3 roughly in the positive direction, while it will give a looser bound than the unitary constraints in the negative direction. In the case of vanishing real parts of $y_L^{tT},y_R^{tT}$, the imaginary parts can only be constrained by unitarity. When the couplings are pure left or pure right, the real parts are also constrained to be less than 3 roughly in the positive direction. For the cases of equal or conjugate $y_L^{tT},y_R^{tT}$, the real parts can be bounded to be less than 1.5 in the positive direction and greater than $-3$ in the negative direction.

As a matter of fact, the behaviours in Fig.~\ref{fig:numerical:ytT} can be explained by the results in Sec.~\ref{subsec:h2gamZ:NP} qualitatively. In $|A_t+A_T+A_{tT}+A_W(\tau_W,\lambda_W)|^2$, $A_{tT}$ can interfere constructively or destructively with $A_W(\tau_W,\lambda_W)$, while $|\mathcal{B}_{tT}|^2$ always enhances the partial width. It will give strong constraints for the constructive case because of double enhancement from $A_{tT},B_{tT}$. $A_{tT}$ is proportional to real parts of $y_{L,R}^{tT}$, while $B_{tT}$ receives the contribution from the imaginary parts of $y_{L,R}^{tT}$. Thus, real parts of $y_{L,R}^{tT}$ are more tightly constrained than the imaginary parts because of the interference with the large $A_W(\tau_W,\lambda_W)$ term. If $A_{tT}>0$ (or $[m_T\mathrm{Re}(y_R^{tT})-(3+2\log r_{tT}^2)m_t\mathrm{Re}(y_L^{tT})]>0$), it will interfere constructively with $A_W(\tau_W,\lambda_W)$. The appearance of $B_{tT}$ will enhance the partial width further; thus, this case is more strongly bounded. If $A_{tT}<0$, there will be some cancellation between the destructive interference with $A_W(\tau_W,\lambda_W)$ and the enhancement from $B_{tT}$. Thus, this case is more loosely bounded.

Although the $m_ty_L^{tT}$ term is suppressed by the factor $\frac{m_t}{m_T}$ compared to the $m_Ty_R^{tT}$ term, it is $\log r_{tT}^2$ enhanced. Thus, we should take both of them into account. Because of $A_{tT}\sim[m_T\mathrm{Re}(y_R^{tT})-(3+2\log r_{tT}^2)m_t\mathrm{Re}(y_L^{tT})]$, the regions of $\mathrm{Re}(y_L^{tT}),\mathrm{Re}(y_R^{tT})$ with same sign are more strongly bounded than those with opposite sign. Because of $\mathcal{B}_{tT}\sim[m_t(1+\log r_{tT}^2)\mathrm{Im}(y_L^{tT})+m_T\mathrm{Im}(y_R^{tT})]$, the regions of $\mathrm{Im}(y_L^{tT}),\mathrm{Im}(y_R^{tT})$ with opposite sign are more strongly bounded than those with same sign (compare $y_L^{tT}=y_R^{tT}$ case with the $y_L^{tT}=(y_R^{tT})^*$ case in Fig.~\ref{fig:numerical:ytT}).

Although the attempts show that the constraints are quite loose, it is still worth investigating the FCNY couplings through the $h\rightarrow\gamma Z$ decay mode. The contributions of FCN couplings are suppressed by both $s_L$ and $\frac{v}{m_T}$. If $s_L$ is not very small, it can give considerable constraints on the FCNY couplings. When $s_L$ becomes very small (say $s_L=0.1$), $h\rightarrow\gamma Z$ decay will lose the power to constrain FCNY couplings (looser than the perturbative unitarity bound). When $m_T$ becomes very heavy (say TeV), it will also lose the power to constrain FCNY couplings.

In Sec.~\ref{subsec:EDM}, we have illustrated that the top quark EDM may give some bounds on the FCNY couplings. Because we have the identity $y_R^{tT}(y_L^{tT})^*-y_L^{tT}(y_R^{tT})^*=2i(\mathrm{Re}y_L^{tT}\mathrm{Im}y_R^{tT}-\mathrm{Re}y_R^{tT}\mathrm{Im}y_L^{tT})$, the blind directions from top EDM are $y_L^{tT}=0,y_R^{tT}=0,y_L^{tT}=y_R^{tT},\mathrm{Im}y_L^{tT}=\mathrm{Im}y_R^{tT}=0,\mathrm{Re}y_L^{tT}=\mathrm{Re}y_R^{tT}=0$. For the three cases $\mathrm{Im}(y_L^{tT})=\mathrm{Re}(y_R^{tT})=0,\mathrm{Re}(y_L^{tT})=\mathrm{Im}(y_R^{tT})=0,y_L^{tT}=(y_R^{tT})^*$, the top quark EDM can give strong constraints. In Fig.~\ref{fig:numerical:ytT}, we also show the allowed regions from top quark EDM at $90\%$ CL and perturbative unitarity (yellow) for these three scenarios. From these plots, we can find that the off-axis regions are strongly bounded by top EDM, while it loses the constraining power in the near axis regions.

By the way, $h\rightarrow\gamma\gamma$ depends only on the same flavour Yukawa couplings, while $h\rightarrow\gamma Z$ decay is also controlled by the FCN couplings. By combing $h\rightarrow\gamma\gamma,\gamma Z$ together, it is possible to disentangle the FCNY couplings from the same flavour Yukawa couplings. For the doublet and triplets VLQ cases, there are extra heavy quarks besides the $T_{L,R}$. The new heavy quarks can contribute to the $h\rightarrow\gamma Z$ decay; thus, the FCNY coupling constraints will be quite loose. Certainly, the FCN couplings can show up in other processes, too. For example, we can search for new physics through the di-Higgs production \cite{Plehn:1996wb, Lu:2015jza, Cao:2016zob}, while the $gg\rightarrow hh$ process suffers from the anomalous $hhh$ coupling. The $e^+e^-\rightarrow h\gamma$ production at electron-positron colliders is also an interesting process and it has drawn much attention of the community. It can also be a probe of the anomalous $h\gamma Z$ and $h\gamma\gamma$ couplings. The SM analysis for this process is given in Refs.~\cite{Barroso:1985et, Abbasabadi:1995rc, Djouadi:1996ws}. There are also some works on this process in many new physics models, for example, the MSSM \cite{Djouadi:1996ws, Gounaris:2015tna, Heinemeyer:2015qbu, Demirci:2019ush}, extended scalar sector models \cite{Arhrib:2014pva, Kanemura:2018esc}, VLQ models \cite{Raissi:2018bgt}, effective field theory framework \cite{Gounaris:1995mx, Ren:2015uka, Cao:2015iua, Dedes:2019bew}, and simplified scenarios \cite{Li:2015kxc}. Besides, we can also probe the FCNY couplings through direct production processes $pp\rightarrow Tth,Tt,ThW,Thj$. But they suffer from a low event rate. Although the FCNY couplings may also be constrained from other processes, the detailed analyses in these channels are beyond the scope of this work.
\section{Summary and conclusions}\label{sec:summary}
There can exist FCN interactions between the top quark and new heavy quark. To unravel the nature of flavour structure and EWSB, it is of great importance to probe such couplings. Unfortunately, it is difficult to constrain the FCN couplings at both current and future experiments. Here, we show how to bound the FCNY couplings in simplified singlet $T_L,T_R$ extended models generally.

In this paper, we have summarized the main constraints from theoretical and experimental viewpoints. By turning off other couplings naively, we get the perturbative unitarity bounds on $|y_{L,R}^{tT}|$. After considering the constraints from direct search, $S,T$ parameters, top physics, and Higgs signal strength, we take $m_T=400$~GeV and $s_L=0.2$ as the benchmark point to get the optimal situation. Under this benchmark point, we consider the future bounds from $h\rightarrow\gamma Z$ decay at HL-LHC numerically. The real parts of $y_{L,R}^{tT}$ in the positive direction can be limited to be less than $1.5\sim3$ because of the double enhancement. For the real parts of $y_{L,R}^{tT}$ in the negative direction, they are mainly bounded by the perturbative unitarity. Finally, we find that top quark EDM can give stronger bounds (especially the imaginary parts of $y_{L,R}^{tT}$) than the perturbative unitarity and $h\rightarrow\gamma Z$ decay in the off-axis regions for some scenarios.
\begin{acknowledgements}
We would like to thank Gang Li, Zhao Li, Ying-nan Mao, Cen Zhang, and Hao Zhang for helpful discussions. We also thank Jordy de Vries for directing our attention to the latest constraints on top quark EDM. 
\end{acknowledgements}
\bibliographystyle{unsrt}
\bibliography{H2gamZ}
\section*{Appendix}
\begin{appendices}
\section{Perturbative unitarity analysis}\label{app:unitary}
\subsection{Two-fermion process analysis}
For the two-fermion process, we take the $t\bar{t}\rightarrow hh$ process as an example.
\begin{figure}[!h]
\includegraphics[scale=0.45]{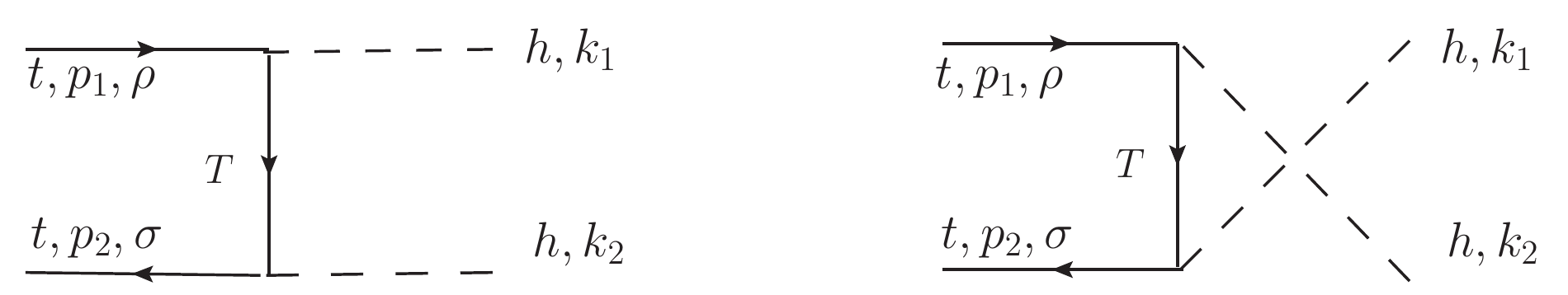}
\caption{Feynman diagrams for the $t\bar{t}\rightarrow hh$ scattering process.}\label{fig:tt2hh}
\end{figure}
In Fig.~\ref{fig:tt2hh}, we give the Feynman diagrams \footnote{The diagrams mediated by $s$-channel Higgs propagator vanish in the high energy limit because of the $\frac{1}{(p_1+p_2)^2-m_h^2}$ suppression.}. The amplitude with general helicity can be written as
\begin{align}
&i\mathcal{M}^{\rho\sigma}(t\bar{t}\rightarrow hh)\nonumber\\
=&-i\bar{v}^{\sigma}(p_2)(y_L^{tT}\omega_-+y_R^{tT}\omega_+)(\frac{1}{\slash\!\!\!{p_1}-\slash\!\!\!{k_1}-m_T}+\frac{1}{\slash\!\!\!{p_1}-\slash\!\!\!{k_2}-m_T})[(y_R^{tT})^*\omega_-+(y_L^{tT})^*\omega_+]u^{\rho}(p_1)\nonumber\\
=&-i\bar{v}^{\sigma}(p_2)[\frac{m_T(|y_L^{tT}|^2\omega_-+|y_R^{tT}|^2\omega_+)+m_T(y_L^{tT}(y_R^{tT})^*\omega_-+y_R^{tT}(y_L^{tT})^*\omega_+)-(|y_L^{tT}|^2\omega_-+|y_R^{tT}|^2\omega_+)\slash\!\!\!{k_1}}{(p_1-k_1)^2-m_T^2}\nonumber\\
&+\frac{m_T(|y_L^{tT}|^2\omega_-+|y_R^{tT}|^2\omega_+)+m_T(y_L^{tT}(y_R^{tT})^*\omega_-+y_R^{tT}(y_L^{tT})^*\omega_+)-(|y_L^{tT}|^2\omega_-+|y_R^{tT}|^2\omega_+)\slash\!\!\!{k_2}}{(p_1-k_2)^2-m_T^2}]u^{\rho}(p_1).
\end{align}
In the high energy limit $p_{1,2}^0\rightarrow\infty$, it can be approximated as
\begin{align}
&i\mathcal{M}^{\rho\sigma}(t\bar{t}\rightarrow hh)\approx i\bar{v}^{\sigma}(p_2)(|y_L^{tT}|^2\omega_-+|y_R^{tT}|^2\omega_+)[\frac{\slash\!\!\!{k_1}}{(p_1-k_1)^2-m_T^2}+\frac{\slash\!\!\!{k_2}}{(p_1-k_2)^2-m_T^2}]u^{\rho}(p_1).
\end{align}
To calculate the above amplitude, we need to choose a reference frame. In the center of mass (COM) frame of initial particles, we can parametrize the momenta $p_1,p_2,k_1,k_2$ and spinors as follows \cite{Maltoni:2001dc, Denner:1991kt}:
\begin{align*}
&p_1^{\mu}=(E_p,0,0,|\vec{p}|),\quad p_2^{\mu}=(E_p,0,0,-|\vec{p}|),\nonumber\\
&k_1^{\mu}=(E_k,|\vec{k}|\sin\theta,0,|\vec{k}|\cos\theta),\quad k_2^{\mu}=(E_k,-|\vec{k}|\sin\theta,0,-|\vec{k}|\cos\theta),\nonumber\\
&s=(p_1+p_2)^2=(2E_p)^2,\quad t=(p_1-k_1)^2,\quad u=(p_1-k_2)^2,
\end{align*}
\begin{align}
&u^+(p_1)=\left[\begin{array}{c}
	\sqrt{E_p-|\vec{p}|}\xi^+ \\ \sqrt{E_p+|\vec{p}|}\xi^+
	\end{array} \right],~
 u^-(p_1)=\left[\begin{array}{c}
	\sqrt{E_p+|\vec{p}|}\xi^- \\ \sqrt{E_p-|\vec{p}|}\xi^-
	\end{array} \right],~
 \xi^+=\left[\begin{array}{c}1\\0\end{array}\right],~
 \xi^-=\left[\begin{array}{c}0\\1\end{array}\right],\nonumber\\
&v^+(p_2)=\left[\begin{array}{c}
	\sqrt{E_p+|\vec{p}|}\eta^+ \\ -\sqrt{E_p-|\vec{p}|}\eta^+
	\end{array} \right],~
 v^-(p_2)=\left[\begin{array}{c}
	\sqrt{E_p-|\vec{p}|}\eta^- \\ -\sqrt{E_p+|\vec{p}|}\eta^-
	\end{array} \right],~
 \eta^+=\left[\begin{array}{c}-1\\0\end{array}\right],~
 \eta^-=\left[\begin{array}{c}0\\-1\end{array}\right].
\end{align}
In the high energy limit, we have:
\begin{align*}
&p_1^{\mu}\approx(E,0,0,E),\quad p_2^{\mu}\approx(E,0,0,-E),\nonumber\\
&k_1^{\mu}\approx(E,E\sin\theta,0,E\cos\theta),\quad k_2^{\mu}\approx(E,-E\sin\theta,0,-E\cos\theta),\nonumber\\
&s\approx(2E)^2,\quad t\approx-2E^2(1-\cos\theta),\quad u\approx-2E^2(1+\cos\theta).
\end{align*}
\begin{align}
&u^+(p_1)\approx\sqrt{2E}\left[\begin{array}{c}
	\vec{0} \\ \xi^+
	\end{array} \right],~
 u^-(p_1)\approx\sqrt{2E}\left[\begin{array}{c}
	\xi^- \\ \vec{0}
	\end{array} \right],~
 v^+(p_2)\approx\sqrt{2E}\left[\begin{array}{c}
	\eta^+ \\ \vec{0}
	\end{array} \right],~
 v^-(p_2)\approx\sqrt{2E}\left[\begin{array}{c}
	\vec{0} \\ -\eta^-
	\end{array} \right].
\end{align}
Thus, we derive the following results:
\begin{align*}
&i\mathcal{M}^{++}(t\bar{t}\rightarrow hh)\nonumber\\
\approx&-i|y_R^{tT}|^2[\frac{1}{(m_h^2+m_t^2-m_T^2)/(2E^2)-(1-\cos\theta)}-\frac{1}{(m_h^2+m_t^2-m_T^2)/(2E^2)-(1+\cos\theta)}],\nonumber\\
&i\mathcal{M}^{+-}(t\bar{t}\rightarrow hh)\nonumber\\
\approx&-i|y_L^{tT}|^2\sin\theta[\frac{1}{(m_h^2+m_t^2-m_T^2)/(2E^2)-(1-\cos\theta)}-\frac{1}{(m_h^2+m_t^2-m_T^2)/(2E^2)-(1+\cos\theta)}],
\end{align*}
\begin{align}
&i\mathcal{M}^{-+}(t\bar{t}\rightarrow hh)\nonumber\\
\approx&-i|y_R^{tT}|^2\sin\theta[\frac{1}{(m_h^2+m_t^2-m_T^2)/(2E^2)-(1-\cos\theta)}-\frac{1}{(m_h^2+m_t^2-m_T^2)/(2E^2)-(1+\cos\theta)}],\nonumber\\
&i\mathcal{M}^{--}(t\bar{t}\rightarrow hh)\nonumber\\
\approx&i|y_L^{tT}|^2[\frac{1}{(m_h^2+m_t^2-m_T^2)/(2E^2)-(1-\cos\theta)}-\frac{1}{(m_h^2+m_t^2-m_T^2)/(2E^2)-(1+\cos\theta)}].
\end{align}
As we can see, there is no $S$-wave in this channel, namely,
\begin{align}
&a_0^{++}(t\bar{t}\rightarrow hh)\approx a_0^{--}(t\bar{t}\rightarrow hh)\approx a_0^{+-}(t\bar{t}\rightarrow hh)\approx a_0^{-+}(t\bar{t}\rightarrow hh)\approx0.
\end{align}
Of course, there are many other two-fermion processes (for example, $t\bar{T}\rightarrow hh,T\bar{t}\rightarrow hh,T\bar{T}\rightarrow hh,t\bar{t}\rightarrow W^+W^-,t\bar{t}\rightarrow ZZ,t\bar{t}\rightarrow Zh$) depending on the initial and final state particles. Actually, all the two-fermion processes do not contribute to the $S$-wave \cite{Chanowitz:1978mv}.
\subsection{Four-fermion process analysis}
For the four-fermion processes, we take the $t\bar{t}\rightarrow T\bar{T}$ process as an example.
\begin{figure}[!h]
\includegraphics[scale=0.45]{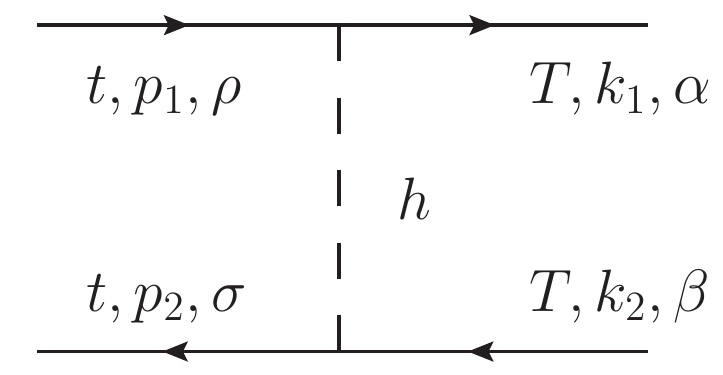}
\caption{Feynman diagram for the $t\bar{t}\rightarrow TT$ scattering process.}\label{fig:tt2TT}
\end{figure}
In Fig.~\ref{fig:tt2TT}, we give the Feynman diagram. The amplitude with general helicity can be written as
\begin{align}
&i\mathcal{M}^{\rho\sigma\alpha\beta}(t\bar{t}\rightarrow T\bar{T})\nonumber\\
=&-\frac{i}{(p_1-k_1)^2-m_h^2}\bar{u}^{\alpha}(k_1)((y_R^{tT})^*\omega_-+(y_L^{tT})^*\omega_+)u^{\rho}(p_1)\bar{v}^{\sigma}(p_2)(y_L^{tT}\omega_-+y_R^{tT}\omega_+)v^{\beta}(k_2).
\end{align}
In the COM frame of initial particles, the representations of spinors are listed as follows:
\begin{align*}
&u^+(p_1)\approx\sqrt{2E}\left[\begin{array}{c}
	\vec{0} \\ \xi^+
	\end{array} \right],~
 u^-(p_1)\approx\sqrt{2E}\left[\begin{array}{c}
	\xi^- \\ \vec{0}
	\end{array} \right],~
 v^+(p_2)\approx\sqrt{2E}\left[\begin{array}{c}
	\eta^+ \\ \vec{0}
	\end{array} \right],~
 v^-(p_2)\approx\sqrt{2E}\left[\begin{array}{c}
	\vec{0} \\ -\eta^-
	\end{array} \right],\nonumber\\
&u^+(k_1)\approx\sqrt{2E}\left[\begin{array}{c}
	\vec{0} \\ \widetilde{\xi}^+
	\end{array} \right],~
 u^-(k_1)\approx\sqrt{2E}\left[\begin{array}{c}
	\widetilde{\xi}^- \\ \vec{0}
	\end{array} \right],~
 v^+(k_2)\approx\sqrt{2E}\left[\begin{array}{c}
	\widetilde{\eta}^+ \\ \vec{0}
	\end{array} \right],~
 v^-(k_2)\approx\sqrt{2E}\left[\begin{array}{c}
	\vec{0} \\ -\widetilde{\eta}^-
	\end{array} \right],
\end{align*}
\begin{align}
&\xi^+=\left[\begin{array}{c}1\\0\end{array}\right],~
\xi^-=\left[\begin{array}{c}0\\1\end{array}\right],~
\eta^+=\left[\begin{array}{c}-1\\0\end{array}\right],~
\eta^-=\left[\begin{array}{c}0\\-1\end{array}\right],~
\gamma^5=\left[\begin{array}{cc}-I_{2\times2}&0_{2\times2}\\0_{2\times2}&I_{2\times2}\end{array}\right],\nonumber\\
&\widetilde{\xi}^+=\left[\begin{array}{c}\cos\frac{\theta}{2}\\ 
	\sin\frac{\theta}{2}\end{array}\right],~
\widetilde{\xi}^-=\left[\begin{array}{c}-\sin\frac{\theta}{2}\\
	\cos\frac{\theta}{2}\end{array}\right],~
\widetilde{\eta}^+=\left[\begin{array}{c}-\cos\frac{\theta}{2}\\ 
	-\sin\frac{\theta}{2}\end{array}\right],~
\widetilde{\eta}^-=\left[\begin{array}{c}\sin\frac{\theta}{2}\\
	-\cos\frac{\theta}{2}\end{array}\right].
\end{align}
Then, we can get the polarized amplitudes:
\begin{align*}
&i\mathcal{M}^{++++}(t\bar{t}\rightarrow T\bar{T})\approx i\mathcal{M}^{+++-}(t\bar{t}\rightarrow T\bar{T})\approx i\mathcal{M}^{++-+}(t\bar{t}\rightarrow T\bar{T})\approx0,\nonumber\\
&i\mathcal{M}^{++--}(t\bar{t}\rightarrow T\bar{T})\approx\frac{isy_R^{tT}(y_L^{tT})^*\sin^2\frac{\theta}{2}}{(p_1-k_1)^2-m_h^2},\nonumber\\
&i\mathcal{M}^{+-++}(t\bar{t}\rightarrow T\bar{T})\approx i\mathcal{M}^{+-+-}(t\bar{t}\rightarrow T\bar{T})\approx i\mathcal{M}^{+---}(t\bar{t}\rightarrow T\bar{T})\approx0,\nonumber\\
&i\mathcal{M}^{+--+}(t\bar{t}\rightarrow T\bar{T})\approx-\frac{is|y_L^{tT}|^2\sin^2\frac{\theta}{2}}{(p_1-k_1)^2-m_h^2},
\end{align*}
\begin{align}
&i\mathcal{M}^{-+++}(t\bar{t}\rightarrow T\bar{T})\approx i\mathcal{M}^{-+-+}(t\bar{t}\rightarrow T\bar{T})\approx i\mathcal{M}^{-+--}(t\bar{t}\rightarrow T\bar{T})\approx0,\nonumber\\
&i\mathcal{M}^{-++-}(t\bar{t}\rightarrow T\bar{T})\approx-\frac{is|y_R^{tT}|^2\sin^2\frac{\theta}{2}}{(p_1-k_1)^2-m_h^2},\nonumber\\
&i\mathcal{M}^{--++}(t\bar{t}\rightarrow T\bar{T})\approx\frac{isy_L^{tT}(y_R^{tT})^*\sin^2\frac{\theta}{2}}{(p_1-k_1)^2-m_h^2},\nonumber\\
&i\mathcal{M}^{--+-}(t\bar{t}\rightarrow T\bar{T})\approx i\mathcal{M}^{---+}(t\bar{t}\rightarrow T\bar{T})\approx i\mathcal{M}^{----}(t\bar{t}\rightarrow T\bar{T})\approx0.
\end{align}
In general, the initial and final states both can be $t\bar{t},t\bar{T},T\bar{t},T\bar{T}$. Thus, the coupled channel matrix is $16\times16$ (four states plus four helicity cases) even if we do not consider the color degrees of freedom. To make the problem as simple as possible, we only turn on the $y_L^{tT},y_R^{tT}$ couplings. Under this consideration, the non-zero coupled channel amplitudes are
\begin{align}
&\mathcal{M}^{++--}(t\bar{t}\rightarrow T\bar{T})\approx\frac{sy_R^{tT}(y_L^{tT})^*\sin^2\frac{\theta}{2}}{(p_1-k_1)^2-m_h^2},\quad \mathcal{M}^{+--+}(t\bar{t}\rightarrow T\bar{T})\approx-\frac{s|y_L^{tT}|^2\sin^2\frac{\theta}{2}}{(p_1-k_1)^2-m_h^2},\nonumber\\
&\mathcal{M}^{-++-}(t\bar{t}\rightarrow T\bar{T})\approx-\frac{s|y_R^{tT}|^2\sin^2\frac{\theta}{2}}{(p_1-k_1)^2-m_h^2},\quad \mathcal{M}^{--++}(t\bar{t}\rightarrow T\bar{T})\approx\frac{sy_L^{tT}(y_R^{tT})^*\sin^2\frac{\theta}{2}}{(p_1-k_1)^2-m_h^2}.
\end{align}
The corresponding $S$-wave amplitudes are calculated to be
\begin{align}
&a_0^{++--}(t\bar{t}\rightarrow T\bar{T})\approx-\frac{y_R^{tT}(y_L^{tT})^*}{16\pi},\quad a_0^{+--+}(t\bar{t}\rightarrow T\bar{T})\approx\frac{|y_L^{tT}|^2}{16\pi},\nonumber\\
&a_0^{-++-}(t\bar{t}\rightarrow T\bar{T})\approx\frac{|y_R^{tT}|^2}{16\pi},\quad a_0^{--++}(t\bar{t}\rightarrow T\bar{T})\approx-\frac{y_L^{tT}(y_R^{tT})^*}{16\pi}.
\end{align}
In the basis of $++,+-,-+,--$, we can get the following coupled channel matrix for this process:
\begin{align}
&a_0(t\bar{t}\rightarrow T\bar{T})=\frac{1}{16\pi}
\left[\begin{array}{cccc}
0&0&0&-y_R^{tT}(y_L^{tT})^*\\
0&0&|y_L^{tT}|^2&0\\
0&|y_R^{tT}|^2&0&0\\
-y_L^{tT}(y_R^{tT})^*&0&0&0\\
\end{array}\right].
\end{align}

Similarly, we can get the following coupled channel matrices for the other processes in the basis of $++,+-,-+,--$:
\begin{align}
&a_0(T\bar{T}\rightarrow t\bar{t})=\frac{1}{16\pi}
\left[\begin{array}{cccc}
0&0&0&-y_R^{tT}(y_L^{tT})^*\\
0&0&|y_R^{tT}|^2&0\\
0&|y_L^{tT}|^2&0&0\\
-y_L^{tT}(y_R^{tT})^*&0&0&0\\
\end{array}\right],
\end{align}

\begin{align}
&a_0(t\bar{T}\rightarrow t\bar{T})=\frac{1}{16\pi}
\left[\begin{array}{cccc}
-|y_L^{tT}|^2&0&0&y_R^{tT}(y_L^{tT})^*\\
0&0&0&0\\
0&0&0&0\\
y_L^{tT}(y_R^{tT})^*&0&0&-|y_R^{tT}|^2\\
\end{array}\right],
\end{align}

\begin{align}
&a_0(t\bar{T}\rightarrow T\bar{t})=\frac{1}{16\pi}
\left[\begin{array}{cccc}
-(y_L^{tT})^*(y_R^{tT})^*&0&0&0\\
0&0&(y_L^{tT})^*(y_R^{tT})^*&0\\
0&(y_L^{tT})^*(y_R^{tT})^*&0&0\\
0&0&0&-(y_L^{tT})^*(y_R^{tT})^*\\
\end{array}\right],
\end{align}

\begin{align}
&a_0(T\bar{t}\rightarrow t\bar{T})=\frac{1}{16\pi}
\left[\begin{array}{cccc}
-y_L^{tT}y_R^{tT}&0&0&0\\
0&0&y_L^{tT}y_R^{tT}&0\\
0&y_L^{tT}y_R^{tT}&0&0\\
0&0&0&-y_L^{tT}y_R^{tT}\\
\end{array}\right],
\end{align}

\begin{align}
&a_0(T\bar{t}\rightarrow T\bar{t})=\frac{1}{16\pi}
\left[\begin{array}{cccc}
-|y_R^{tT}|^2&0&0&y_R^{tT}(y_L^{tT})^*\\
0&0&0&0\\
0&0&0&0\\
y_L^{tT}(y_R^{tT})^*&0&0&-|y_L^{tT}|^2\\
\end{array}\right].
\end{align}

In the basis of $t\bar{t},T\bar{T},t\bar{T},T\bar{t}$, we can get the following coupled channel matrix for all the four-fermion processes without regard to the quark color:
\begin{align}
a_0=\left[\begin{array}{cccc}
0_{4\times4}&a_0(t\bar{t}\rightarrow T\bar{T})&0_{4\times4}&0_{4\times4}\\
a_0(T\bar{T}\rightarrow t\bar{t})&0_{4\times4}&0_{4\times4}&0_{4\times4}\\
0_{4\times4}&0_{4\times4}&a_0(t\bar{T}\rightarrow t\bar{T})&a_0(t\bar{T}\rightarrow T\bar{t})\\
0_{4\times4}&0_{4\times4}&a_0(T\bar{t}\rightarrow t\bar{T})&a_0(T\bar{t}\rightarrow T\bar{t})\\
\end{array}\right].
\end{align}
In the above, we write the coupled channel matrix in the block form. Obviously, this square matrix is $16\times16$.
Then we can get the eigenvalues of $a_0$ as follows \footnote{When we take the quark color into account further, the matrix will become $48\times48$. Although the unitarity bounds may be improved, the matrix will be quite large and complex to deal with.}:
\begin{align}
&\lambda_1^+=\frac{1}{16\pi}|y_L^{tT}|^2,\quad \lambda_1^-=-\frac{1}{16\pi}|y_L^{tT}|^2,\quad \lambda_2^+=\frac{1}{16\pi}|y_R^{tT}|^2,\quad \lambda_2^-=-\frac{1}{16\pi}|y_R^{tT}|^2,\nonumber\\
&\lambda_3^+=\frac{1}{16\pi}|y_L^{tT}||y_R^{tT}|,\quad \lambda_3^-=-\frac{1}{16\pi}|y_L^{tT}||y_R^{tT}|,\nonumber\\
&\lambda_4^+=\frac{1}{32\pi}(\sqrt{(|y_L^{tT}|^2+|y_R^{tT}|^2)^2+12|y_L^{tT}|^2|y_R^{tT}|^2}-|y_L^{tT}|^2-|y_R^{tT}|^2),\nonumber\\
&\lambda_4^-=\frac{1}{32\pi}(-\sqrt{(|y_L^{tT}|^2+|y_R^{tT}|^2)^2+12|y_L^{tT}|^2|y_R^{tT}|^2}-|y_L^{tT}|^2-|y_R^{tT}|^2),
\end{align}
where $\lambda_1^-,\lambda_2^-$ are doubly degenerate and $\lambda_3^+,\lambda_3^-$ are four-fold degenerate. $S$-wave unitarity requires that all the eigenvalues must satisfy $|\mathrm{Re}(\lambda_i)|\leq\frac{1}{2}$. It will lead to the following constraints \footnote{Remember that the bounds are just a rough estimation. If we turn on the other couplings (say $y_T,\kappa_t$), these constraints may be altered.}:
\begin{align}
\sqrt{(|y_L^{tT}|^2+|y_R^{tT}|^2)^2+12|y_L^{tT}|^2|y_R^{tT}|^2}+|y_L^{tT}|^2+|y_R^{tT}|^2\leq16\pi.
\end{align}
Note that $|y_L^{tT}|\leq\sqrt{8\pi},|y_R^{tT}|\leq\sqrt{8\pi},\sqrt{|y_L^{tT}||y_R^{tT}|}\leq\sqrt{8\pi}$ hold automatically in the above bound.
\section{$h\rightarrow\gamma\gamma$ channel analysis}\label{app:h2gamgam}
\subsection{SM result}
The partial decay width of $h\rightarrow\gamma\gamma$ for SM is given in Refs.~\cite{Gunion:1989we, Djouadi:2005gi}
\begin{align}
\Gamma^{SM}(h\rightarrow\gamma\gamma)=\frac{G_F\alpha^2m_h^3}{128\sqrt{2}\pi^3}|\sum_fN_f^CQ_f^2F_f(\tau_f)+F_W(\tau_W)|^2~(\tau_f=\frac{4m_f^2}{m_h^2},\tau_W=\frac{4m_W^2}{m_h^2}),
\end{align}
with the $F_f,F_W$ defined by
\begin{align}
&F_f(\tau_f)\equiv-2\tau_f[1+(1-\tau_f)f(\tau_f)],\quad F_W(\tau_W)\equiv2+3\tau_W+3\tau_W(2-\tau_W)f(\tau_W),\nonumber\\
&f(\tau)\equiv
              \begin{cases}
              \arcsin^2(\frac{1}{\sqrt{\tau}}), &\mathrm{for}~\tau\geq1\\
              -\frac{1}{4}[\log\frac{1+\sqrt{1-\tau}}{1-\sqrt{1-\tau}}-i\pi]^2, &\mathrm{for}~\tau<1
              \end{cases}.
\end{align}
For the fermionic part, the top quark is dominated because of the largest Yukawa coupling. Numerically, we can get $N_t^CQ_t^2F_f(\tau_t)\sim-1.84,F_W(\tau_W)\sim8.32$. This means the gauge boson contributions are almost 4.5 times larger than the fermionic ones.
\subsection{New physics result}
Due to $U_{EM}(1)$ gauge symmetry, the $h\rightarrow\gamma\gamma$ amplitude possesses the following tensor structure:
\begin{align}
&i\mathcal{M}=i\epsilon_{\mu}(p_1)\epsilon_{\nu}(p_2)(p_2^{\mu}p_1^{\nu}-p_1\cdot p_2g^{\mu\nu})\mathcal{A}^\gamma,\qquad\mathcal{A}^\gamma\equiv\frac{e^2}{8\pi^2v}(-F_W(\tau_W)+\mathcal{A}_t^\gamma+\mathcal{A}_T^\gamma).
\end{align}
The expressions of $\mathcal{A}_t^\gamma,\mathcal{A}_T^\gamma$ are given as
\begin{align}
&\mathcal{A}_t^\gamma=-N_t^CQ_t^2\kappa_tF_f(\tau_t),\quad \mathcal{A}_T^\gamma=N_T^CQ_T^2\frac{y_{T}v}{m_T}F_f(\tau_T).
\end{align}
Taking the mass of $t,T$ quarks to be infinity, they can be expanded as
\begin{align}
&\mathcal{A}_t^\gamma\approx-N_t^CQ_t^2\kappa_t(-\frac{4}{3}-\frac{7m_h^2}{90m_t^2}),\quad \mathcal{A}_T^\gamma\approx N_T^CQ_T^2\frac{y_{T}v}{m_T}(-\frac{4}{3}-\frac{7m_h^2}{90m_T^2}),\nonumber\\
&\mathcal{A}_t^\gamma+\mathcal{A}_T^\gamma\approx\frac{4}{3}N_t^CQ_t^2[(\kappa_t-\frac{y_{T}v}{m_T})+\kappa_t\cdot\frac{7m_h^2}{120m_t^2}-\frac{y_{T}v}{m_T}\cdot\frac{7m_h^2}{120m_T^2}].
\end{align}
The partial decay width formula is computed as \footnote{The $h\rightarrow gg$ partial decay width is similar to the fermionic part of the $\gamma\gamma$ decay.}:
\begin{align}
&\Gamma(h\rightarrow\gamma\gamma)=\frac{G_F\alpha^2m_h^3}{128\sqrt{2}\pi^3}|A_t^\gamma+A_T^\gamma-F_W(\tau_W)|^2=\frac{G_F\alpha^2m_h^3}{128\sqrt{2}\pi^3}
|N_t^CQ_t^2[\kappa_tF_f(\tau_t)-\frac{y_{T}v}{m_T}F_f(\tau_T)]
+F_W(\tau_W)|^2.
\end{align}

In Tab.~\ref{tab:gamgam:AB}, we list the expressions of $A_t^\gamma+A_T^\gamma$ in several models, where we have neglected the $\frac{1}{m_{t,T}^4}$ suppressed terms. We can see that the $A_t^\gamma+A_T^\gamma$ in VLQT and VLQT+S models are close to those in SM. In fact, it is difficult to detect VLQ in the $h\gamma\gamma$ decay channel \cite{Aguilar-Saavedra:2013qpa}.
\begin{sloppypar}
\begin{table}[!h]
\begin{tabular}{c|c}
\hline
\diagbox{\qquad}{\qquad}& $\bar{A}_t^\gamma+\bar{A}_T^\gamma$ \\
\hline
General & $\frac{4}{3}[(\kappa_t-\frac{y_{T}v}{m_T})+\kappa_t\cdot\frac{7m_h^2}{120m_t^2}-\frac{y_{T}v}{m_T}\cdot\frac{7m_h^2}{120m_T^2}]$ \\
\hline
SM & $\frac{4}{3}(1+\frac{7m_h^2}{120m_t^2})$ \\
\hline
VLQT & $\frac{4}{3}(1+c_L^2\cdot\frac{7m_h^2}{120m_t^2}+s_L^2\cdot\frac{7m_h^2}{120m_T^2})$ \\
\hline
VLQT+S & \makecell{$\frac{4}{3}[c_{\theta}-\frac{v}{m_T}\mathrm{Re}(y_T^S)s_{\theta}\frac{c_R}{c_L}+(c_L^2c_{\theta}-\frac{v}{m_T}\mathrm{Re}(y_T^S)s_{\theta}\frac{s_L^2c_R}{c_L})\cdot\frac{7m_h^2}{120m_t^2}$\\$+(s_L^2c_{\theta}-\frac{v}{m_T}\mathrm{Re}(y_T^S)s_{\theta}c_Lc_R)\cdot\frac{7m_h^2}{120m_T^2}]$} \\
\hline
\end{tabular}
\caption{The expressions of $\bar{A}_t^\gamma+\bar{A}_T^\gamma$ in the SM, VLQT, and VLQT+S under the heavy quark limit. Here, we extract the common factor $N_t^CQ_t^2$ for convenience, that is, redefinition of $A^\gamma$ with $N_t^CQ_t^2\bar{A}^\gamma$. We take $\bar{A}_T^\gamma=0$ naively in the SM because of the absence of a $T$ quark.}\label{tab:gamgam:AB}
\end{table}
\end{sloppypar}

\section{Asymptotic behaviors of the loop functions}\label{app:loopfuns}
The $B_0$ function is defined as \cite{Denner:1991kt}
\begin{align}
&B_0(k^2,m_0^2,m_1^2)\nonumber\\
\equiv&\frac{(2\pi\mu)^{4-D}}{i\pi^2}\int d^Dq\frac{1}{(q^2-m_0^2)[(q+k)^2-m_1^2]}\nonumber\\
=&\Delta_{\epsilon}-\int_0^1dx\log\frac{xm_1^2+(1-x)m_0^2-x(1-x)k^2}{\mu^2}~(\Delta_{\epsilon}=\frac{1}{\epsilon}-\gamma_E+\log4\pi,~D=4-2\epsilon).
\end{align}
In the limit of $k^2\ll m_0^2,m_1^2$, the $B_0$ function can be expanded as
\begin{align}
&B_0(k^2,m_0^2,m_1^2)\nonumber\\
=&B_0(0,m_0^2,m_1^2)+\frac{\partial B_0(k^2,m_0^2,m_1^2)}{\partial k^2}|_{k^2=0}k^2+\mathcal{O}(\frac{k^4}{m_0^4},\frac{k^4}{m_0^2m_1^2},\frac{k^4}{m_1^4})\nonumber\\
=&\Delta_{\epsilon}+1-\frac{m_0^2\log\frac{m_0^2}{\mu^2}-m_1^2\log\frac{m_1^2}{\mu^2}}{m_0^2-m_1^2}+\frac{m_0^4-m_1^4+2m_0^2m_1^2\log\frac{m_1^2}{m_0^2}}{2(m_0^2-m_1^2)^3}k^2+\mathcal{O}(\frac{k^4}{m_0^4},\frac{k^4}{m_0^2m_1^2},\frac{k^4}{m_1^4}).
\end{align}
The $C_0$ function is defined as
\begin{align}
&C_0(k_1^2,k_{12}^2,k_2^2,m_0^2,m_1^2,m_2^2)~(k_{12}\equiv k_1-k_2)\nonumber\\
\equiv&\frac{(2\pi\mu)^{4-D}}{i\pi^2}\int d^Dq\frac{1}{(q^2-m_0^2)[(q+k_1)^2-m_1^2][(q+k_2)^2-m_2^2]}\nonumber\\
=&-\int_0^1\int_0^1\int_0^1dxdydz\frac{\delta(x+y+z-1)}{(yk_1+zk_2)^2+xm_0^2+ym_1^2+zm_2^2-yk_1^2-zk_2^2}.
\end{align}
Then, we have
\begin{align}
&C_0(0,m_Z^2,m_h^2,m_t^2,m_t^2,m_T^2)=C_0(0,m_Z^2,m_h^2,m_T^2,m_t^2,m_t^2)\nonumber\\
=&-\int_0^1\int_0^1\int_0^1dxdydz\frac{\delta(x+y+z-1)}{[yp_1+z(p_1+p_2)]^2+(x+y)m_t^2+zm_T^2-yp_1^2-z(p_1+p_2)^2}\nonumber\\
=&-\int_0^1\int_0^1\int_0^1dxdydz\frac{\delta(x+y+z-1)}{yz(m_h^2-m_Z^2)+z^2m_h^2+(x+y)m_t^2+zm_T^2-zm_h^2}.
\end{align}
In the limit of $m_h,m_Z\ll m_f$, the $C_0(0,m_Z^2,m_h^2,m_f^2,m_f^2,m_f^2)$ function can be expanded as
\begin{align}
&C_0(0,m_Z^2,m_h^2,m_f^2,m_f^2,m_f^2)=-\frac{1}{2m_f^2}[1+\frac{m_h^2+m_Z^2}{12m_f^2}+\mathcal{O}(\frac{m_h^4,m_h^2m_Z^2,m_Z^4}{m_f^4})].
\end{align}
In the limit of $m_h,m_Z\ll m_t,m_T$, the $C_0(0,m_Z^2,m_h^2,m_t^2,m_t^2,m_T^2)$ function can be expanded as
\begin{align}
&C_0(0,m_Z^2,m_h^2,m_t^2,m_t^2,m_T^2)=C_0(0,0,0,m_t^2,m_t^2,m_T^2)+\frac{\partial C_0(0,m_Z^2,m_h^2,m_t^2,m_t^2,m_T^2)}{\partial m_h^2}|_{(m_h=0,m_Z=0)}m_h^2\nonumber\\
&+\frac{\partial C_0(0,m_Z^2,m_h^2,m_t^2,m_t^2,m_T^2)}{\partial m_Z^2}|_{(m_h=0,m_Z=0)}m_Z^2+\mathcal{O}(\frac{m_h^4,m_h^2m_Z^2,m_Z^4}{m_t^6,m_t^4m_T^2,m_t^2m_T^4,m_T^6}),
\end{align}
with
\begin{align}
&C_0(0,0,0,m_t^2,m_t^2,m_T^2)=C_0(0,0,0,m_T^2,m_t^2,m_t^2)=\frac{1}{m_T^2}\frac{1-r_{tT}^2+\log r_{tT}^2}{(1-r_{tT}^2)^2}~(r_{tT}=\frac{m_t}{m_T}),\nonumber\\
&\frac{\partial C_0(0,m_Z^2,m_h^2,m_t^2,m_t^2,m_T^2)}{\partial m_h^2}|_{(m_h=0,m_Z=0)}=\frac{\partial C_0(0,m_Z^2,m_h^2,m_t^2,m_t^2,m_T^2)}{\partial m_Z^2}|_{(m_h=0,m_Z=0)}\nonumber\\
=&\frac{1}{m_T^4}\frac{2(1+2r_{tT}^2)\log r_{tT}^2+5-4r_{tT}^2-r_{tT}^4}{4(1-r_{tT}^2)^4}.
\end{align}
Thus, we have
\begin{align}
&C_0(0,m_Z^2,m_h^2,m_t^2,m_t^2,m_T^2)\approx\frac{1}{m_T^2}[1+\log r_{tT}^2+\frac{m_h^2+m_Z^2}{m_T^2}\frac{5+2(1+2r_{tT}^2)\log r_{tT}^2}{4}].
\end{align}
For the case of $C_0(0,m_Z^2,m_h^2,m_T^2,m_T^2,m_t^2)$, we can get the corresponding results via the replacement $m_t\leftrightarrow m_T$. For example, we have
\begin{align}
&C_0(0,m_Z^2,m_h^2,m_T^2,m_T^2,m_t^2)\approx-\frac{1}{m_T^2}[1+r_{tT}^2\log r_{tT}^2+\frac{m_h^2+m_Z^2}{m_T^2}\frac{1+4r_{tT}^2\log r_{tT}^2}{4}].
\end{align}
Here, we also give the heavy $m_f$ expansion of the following functions:\\
\begin{align}
&f(\tau_f)=\frac{m_h^2}{4m_f^2}+\frac{m_h^4}{48m_f^4}+\mathcal{O}(\frac{m_h^6}{m_f^6}),\quad F_f(\tau_f)=-\frac{4}{3}-\frac{7m_h^2}{90m_f^2}+\mathcal{O}(\frac{m_h^4}{m_f^4}).
\end{align}
and
\begin{align}
A_f(\tau_f,\lambda_f)=&\frac{m_f^2}{m_h^2-m_Z^2}[(m_h^2-m_Z^2-4m_f^2)C_0(0,m_Z^2,m_h^2,m_f^2,m_f^2,m_f^2)-2m_Z^2\frac{B_0(m_h^2,m_f^2,m_f^2)-B_0(m_Z^2,m_f^2,m_f^2)}{m_h^2-m_Z^2}-2]\nonumber\\
=&-\frac{1}{3}-\frac{7m_h^2+11m_Z^2}{360m_f^2}+\mathcal{O}(\frac{m_h^4,m_h^2m_Z^2,m_Z^4}{m_f^4}).
\end{align}
In the following, we list some special limits of the loop integrals:
\begin{align}
&\lim_{\tau_f\rightarrow0}F_f(\tau_f)=0,\quad \lim_{\tau_f\rightarrow\infty}F_f(\tau_f)=-\frac{4}{3},\nonumber\\
&\lim_{\tau_W\rightarrow0}F_W(\tau_W)=2,\quad \lim_{\tau_W\rightarrow\infty}F_W(\tau_W)=7,\nonumber\\
&\lim_{\lambda\rightarrow\infty}I_1(\tau,\lambda)=\frac{\tau^2}{2}f(\tau)-\frac{\tau}{2},\quad \lim_{\lambda\rightarrow\infty}I_2(\tau,\lambda)=\frac{\tau}{2}f(\tau),\nonumber\\
&\lim_{\lambda_f\rightarrow\infty}A_f(\tau_f,\lambda_f)=\frac{\tau_f}{2}[(\tau_f-1)f(\tau_f)-1]=\frac{1}{4}F_f(\tau_f),\nonumber\\
&\lim_{\lambda_W\rightarrow\infty}A_W(\tau_W,\lambda_W)=\frac{1}{2t_W}[(5-t_W^2)\tau_W(2-\tau_W)f(\tau_W)+2+5\tau_W-t_W^2(2+\tau_W)].
\end{align}
\end{appendices}
\clearpage
\newpage
\end{document}